\documentclass[a4paper,11pt]{article}
\pdfoutput=1 

\usepackage{jheppub} 

\usepackage[T1]{fontenc} 
\usepackage{empheq,mathtools}
\usepackage{tikz}
\usetikzlibrary{arrows.meta, patterns, decorations.pathmorphing, positioning}
\usepackage{array}

\usepackage{braket} 
\usepackage{tabularray}
\usepackage{makecell}

\title{\boldmath Holographic thermal correlators from recursions}

\author{Jie Ren and Zhe Yu}

\affiliation{School of Physics, Sun Yat-sen University, Guangzhou, 510275, China}

\emailAdd{renjie7@mail.sysu.edu.cn, yuzh53@mail2.sysu.edu.cn}

\abstract{We express holographic thermal correlators using a recurrence relation of $\{a_n\}$ at $n\to\infty$, building on recent advances in the connection formula for the Heun equation. We consider two gravitational solutions that correspond to distinct states in different subsectors of $\mathcal{N}=4$ super-Yang-Mills theory at finite temperature and density. The first is the Reissner-Nordstr\"{o}m-AdS$_5$ black hole, which has finite entropy at zero temperature, and the second is a charged dilatonic black hole in AdS$_5$, which has zero entropy at zero temperature. In both cases, we perturb the system with a charged scalar field and express the perturbation equation in terms of the Heun equation. We find interesting moving patterns of the poles of the correlators including eigenvalue repulsions. We discuss the relation between the recurrence relation and the Virasoro conformal block as two equivalent approaches to write the connection formula for the Heun equation.} 

\arxivnumber{2412.02608}

\begin{document} 
\maketitle
\flushbottom

\section{Introduction}
\label{sec:intro}

Holographic correlators are calculated in terms of the Gubser-Klebanov-Polyakov-Witten (GKPW) prescription, which provides a bridge between bulk gravitational physics in AdS space and correlation functions in the boundary CFT \cite{Maldacena:1997re,Gubser:1998bc,Witten:1998qj,Son:2002sd}. A crucial step for calculating holographic thermal correlators is to solve the perturbation equations of a black hole, which often reduce to a second-order ordinary differential equation (ODE) with boundary conditions imposed at two of its multiple singularities. As the holographic correlators involve the connection between the black hole horizon and the AdS boundary, we need to obtain the connection formula of the independent solutions around two different singularities where the boundary conditions are imposed. When the perturbation equation has four or more singular points, holographic thermal correlators are not available explicitly.

The key to solving the holographic correlators is a connection problem, for which the perturbation equation can often be transformed into the form \cite{Lisovyy:2022flm}
\begin{equation}
\left(\frac{d^{2}}{dz^{2}} +\frac{\frac{1}{4} -\theta _{0}^{2}}{z^{2}} +\frac{\frac{1}{4} -\theta _{1}^{2}}{(z-1)^{2}} +\frac{U(z)}{z(z-1)}\right) \psi (z)=0, \label{eq:gen2ndODE}
\end{equation}
where $U(z)$ is analytic inside the disk $|z|<R$ ($R>1$). The linearly independent solutions at $z=0$ and $z=1$ are
\begin{align}
    \psi^{[0]}_{\pm}(z)=&\,z^{\frac12\mp\theta_0}\biggl[1+\sum_{k=1}^\infty \varphi^{[0]}_{\pm,k} z^k\biggr],& \text{as }z\to0^+,\label{eq:psi0pm}\\ 
   	\psi^{[1]}_{\pm}(z)=&\,(1-z)^{\frac12\mp\theta_1}\biggl[1+\sum_{k=1}^\infty \varphi^{[1]}_{\pm,k}(1-z)^k\biggr],& \text{as }z\to1^-,\label{eq:psi1pm}
\end{align}
where $1/2\mp\theta_0$ are the two exponents at $z=0$, and $1/2\mp\theta_1$ are the two exponents at $z=1$. We first assume $\theta_0, \theta_1\notin\mathbb{Z}/2$. The two linearly independent solutions at $z=0$ and $z=1$ are related by
\begin{equation}
\setlength\arraycolsep{5pt}
    \begin{pmatrix}
         \psi_1^{[0]}(z)\\
         \psi_2^{[0]}(z) 
    \end{pmatrix}=
    \begin{pmatrix}
         \mathsf{C}_{11} & \mathsf{C}_{12}\\
         \mathsf{C}_{21} & \mathsf{C}_{22}
    \end{pmatrix}
    \begin{pmatrix}
         \psi_1^{[1]}(z)\\
         \psi_2^{[1]}(z) 
    \end{pmatrix},
\label{eq:connCij}
\end{equation}
where the connection matrix $\mathsf{C}_{ij}$ depends on the parameters of the equation. As solutions to \eqref{eq:gen2ndODE}, $\psi _{+}^{[0]}$ and $\psi _{-}^{[0]}$ are related by $\theta _{0}\rightarrow -\theta_{0}$, and $\psi _{+}^{[1]}$ and $\psi _{-}^{[1]}$ are related by $\theta _{1}\rightarrow -\theta_{1}$. Thus, \eqref{eq:connCij} can be rewritten as
\begin{equation}
\psi _{\epsilon }^{[0]} (z)=\sum _{\epsilon ^{\prime } =\pm }\mathsf{C}_{\epsilon \epsilon ^{\prime }}( \theta _{0} ,\theta _{1}) \psi _{\epsilon ^{\prime }}^{[1]} (z)=\sum _{\epsilon ^{\prime } =\pm }\mathsf{C}\left( \epsilon \theta _{0} ,\epsilon ^{\prime } \theta _{1}\right) \psi _{\epsilon ^{\prime }}^{[1]}(z)\,,\label{eq:psiconn}
\end{equation}
where the connection matrix is determined by a single function. In practice, assuming that $\psi_1^{[0]}$ satisfies the infalling boundary condition at the horizon, and $\{\psi_1^{[1]}, \psi_2^{[1]}\}$ are leading and subleading behaviors near the AdS boundary, the retarded Green's function is given by $G=\mathsf{C}_{12}/\mathsf{C}_{11}$ up to a prefactor. When $\theta_1\in\mathbb{Z}/2$, there will be logarithmic terms. In particular, the quasinormal modes are determined by $\mathsf{C}_{11}=0$.

The purpose of this paper is to express the holographic thermal correlators in terms of simply a recurrence relation, building on recent advances \cite{Lisovyy:2022flm} in mathematics of the connection formula for the Heun equation. While this approach is equivalent to employing the semiclassical Virasoro conformal blocks to express the connection formula (the Trieste formula \cite{Bonelli:2022ten,Bonelli:2021uvf}), it has the advantage that the connection coefficients are determined by a recurrence relation \textit{without referring to any CFT language}. We write the Heun equation in both standard and canonical forms. By expanding perturbatively, the connection coefficients can be analytically obtained order by order. We also give the connection formula when logarithmic terms appear in a Frobenius solution, which is often the case in physics problems.

We apply this method to two gravitational solutions corresponding to different states in $\mathcal{N}=4$ super-Yang-Mills (SYM) theory. One is the Reissner-Nordstr\"{o}m-AdS$_5$ (RN-AdS$_5$) black hole, and the other is the two-charge black hole in AdS$_5$ for the Gubser-Rocha model \cite{Gubser:2009qt}. The two solutions have sharp differences at zero temperature characterized by their IR geometries. The RN-AdS black hole has a finite (nonzero) entropy at zero temperature, and the IR geometry is AdS$_2\times\Sigma^3$. The two-charge black hole in AdS$_5$ has zero entropy at zero temperature, and the IR geometry is conformal to AdS$_2\times\Sigma^3$ \cite{Gubser:2012yb,Ren:2020djc}. At finite temperature, the RN-AdS$_5$ black hole has two horizons, while the two-charge black hole in AdS$_5$ has only one horizon in the spherical and planar cases.

We study the perturbations by a charged scalar field in the background of the RN-AdS$_5$ black hole and the two-charge black hole in AdS$_5$. While the scalar perturbation of the RN-AdS$_5$ black hole has been intensely studied in terms of the Heun equation \cite{Zhang:2020apg,Amado:2021erf,Bhatta:2022wga,Bhatta:2023qcl}, we find that the perturbation equation for the two-charge black hole can also be transformed to a Heun equation with a similar structure. The black hole horizon at finite temperature and the AdS boundary correspond to regular singularities of the perturbation equation. The two-point function of the dual scalar operator is expressed in terms of a recurrence relation. Quasinormal modes are obtained as the poles of the correlators. We find interesting moving patterns of the poles, including the effect of eigenvalue repulsions similar to \cite{Dias:2022oqm,Davey:2023fin}, where quasinormal modes of asymptotic flat black holes were studied.

A significant advancement on the Heun type equations before \cite{Lisovyy:2022flm} is using semiclassical Virasoro conformal blocks, which can be computed in terms of the AGT correspondence \cite{Alday:2009aq}. In this framework, the conformal block is obtained through the Nekrasov–Shatashvili (NS) partition function, which can be expressed as a sum over contributions from supersymmetric instantons \cite{Alba:2010qc, Nekrasov:2013xda}. Pioneering works are \cite{Aminov:2020yma,Bonelli:2022ten,Bonelli:2021uvf}, followed by many applications, such as QNMs of C-metric \cite{Lei:2023mqx}, Kerr-Compton amplitudes \cite{Bautista:2023sdf}, holographic thermal correlators \cite{Dodelson:2022yvn,Bhatta:2022wga,He:2023wcs,Bhatta:2023qcl,Jia:2024zes,BarraganAmado:2024tfu}, \textit{etc}. We review the Trieste formula and specifically illustrate its challenge that some of its parameters are implicitly determined by a transcendental equation involving accessory parameters. While the recurrence relation studied here corresponds to the Trieste formula in the $s$-channel expansion, the recurrence relation corresponding to the $t$-channel expansion has not been found.

This paper is organized as follows. In section~\ref{sec:correlator}, we take two examples of black hole perturbations and show that the perturbation equation can be written in terms of a Heun equation. In section~\ref{sec:Heun}, we apply the perturbative connection formula to black hole perturbations. In section~\ref{sec:Trieste}, we discuss the relation between the two approaches --- conformal block and recurrence relation --- to address the connection problem. Finally, we summarize and propose some open questions. In appendix~\ref{sec:conn}, we review some earlier seminal results on the connection formula. In appendix~\ref{sec:hypergeom}, we give the connection formula for the hypergeometric equation as a comparison. In appendix~\ref{sec:comp}, we discuss the accuracy of the recursion method by comparing it with the pseudospectral method.

\section{Holographic thermal correlators}
\label{sec:correlator}

The perturbation equations of black holes can often be reduced to a second-order ODE, which is classified by its singularities. If there are three regular singularities, the equation can be solved in terms of the hypergeometric function, whose connection formula is well known and reviewed in appendix~\ref{sec:hypergeom}. If there are four regular singularities, the equation can be solved in terms of the Heun function, whose connection formula has been studied in \cite{Bonelli:2022ten,Lisovyy:2022flm} recently. For non-constant curvature spacetimes, we rarely obtain hypergeometric equations. The streamline of calculating the holographic correlators is as follows.
\begin{itemize}
    \item Start with a black hole solution and its perturbations. Here, we consider the case that the perturbation equations can be decoupled and separated into ODEs.
    \item Find an appropriate change of variables to write the equation in the form of \eqref{eq:gen2ndODE} with the horizon at $z=0$ and the AdS boundary at $z=1$.
    \item Write the two linearly independent solutions at $z=0$ and $z=1$ and apply the connection formula.
\end{itemize}
We consider two examples of holographic thermal correlators. The first example has been intensely studied (an early work is \cite{Nunez:2003eq}), while the second example is new. In a broader scope of black hole perturbations, the Heun equation is also applicable. We have attached a Mathematica notebook for practically applying the recurrence relation as an ancillary file.


\subsection{\texorpdfstring{RN-AdS$_5$}{RN-AdS5} black hole}\label{sec:RN}
We take the RN-AdS$_5$ black hole as the background geometry and a charged scalar field $\Phi$ as a perturbation. The action is
\begin{equation}
S=\int d^5x\sqrt{-g}\left(R+\frac{12}{L^2}-\frac{1}{4}F_{\mu\nu}F^{\mu\nu}-|D_\mu\Phi|^2-m^2|\Phi|^2\right),
\end{equation}
where we will set the AdS radius $L=1$. The mass $m$ of the scalar field is related to the scaling dimension $\Delta$ of the scalar operator in the dual CFT by $\Delta(\Delta-4)=m^2L^2$. We assume that $m^2$ is above the Breitenlohner-Freedman (BF) bound: $m^2\geq -4$.

We first write down the background solution when $\Phi=0$. The metric for the spherical, planar, and hyperbolic RN-AdS$_5$ black holes is
\begin{equation}
ds^2=-f(r)dt^2+\frac{1}{f(r)}dr^2+d\Sigma_{3,\kappa}^2,\label{eq:metric-RN}
\end{equation}
where $d\Sigma_{3,\kappa}^2$ is the metric for a three-dimensional unit sphere $\mathbb{S}^3$ ($\kappa=1$), plane $\mathbb{R}^3$ ($\kappa=0$), and hyperbolic space with unit radius $\mathbb{H}^3$ ($\kappa=-1$). The solution to the metric~\eqref{eq:metric-RN} is
\begin{equation}
f(r)=\kappa-\frac{M}{r^2}+\frac{Q^2}{r^4}+r^2=\frac{(r^2-r_+^2)(r^2-r_-^2)(r^2-r_0^2)}{r^4}\,,
\label{eq:metric-RNf}
\end{equation}
where $r_+$ is the radius of the outer horizon, $r_-$ is the radius of the inner horizon, and $r_0$ is a complex ``horizon'' introduced only for simplifying the expressions. The quantities $r_-^2$ and $r_0^2$ are given by
\begin{equation}
r_-^2=\frac{1}{2}\Bigl(-\kappa-r_+^2+\sqrt{(\kappa +r_+^2)^2+4Q^2/r_+^2}\Bigr),\qquad r_0^2=\frac{1}{2}\Bigl(-\kappa-r_+^2-\sqrt{(\kappa +r_+^2)^2+4Q^2/r_+^2}\Bigr).
\end{equation}
We have $r_0^2+r_+^2+r_-^2+\kappa=0$. While the physical parameters are $(M,Q)$, it is more convenient to use $(r_+,Q)$ as the parameter space of the solution. The solution to the gauge field $A=A_tdt$ is
\begin{equation}
A_t(r)=\mu\biggl(1-\frac{r_+^2}{r^2}\biggr),\qquad \mu=\sqrt{3}Q/r_+^2.
\end{equation}

The temperature at the (outer) horizon is
\begin{equation}
T_+=\frac{f'(r_+)}{4\pi}=\frac{1}{2\pi}\frac{(r_+^2-r_-^2)(r_+^2-r_0^2)}{r_+^3}=\frac{1}{2\pi}\biggl(\frac{\kappa}{r_+}+2r_+-\frac{Q^2}{r_+^5}\biggr).
\end{equation}
As we increase the charge $Q$, the temperature decreases. Consequently, the extremal black hole is obtained by
\begin{equation}
    Q_c=r_+^2\sqrt{\kappa +2r_+^2}\,.
\end{equation}
We will also use the temperatures at $r=r_-$ and $r=r_0$ to simplify the expressions later:
\begin{equation}
T_-=\frac{1}{2\pi}\frac{(r_-^2-r_+^2)(r_-^2-r_0^2)}{r_-^3},\qquad T_0=\frac{1}{2\pi}\frac{(r_0^2-r_+^2)(r_0^2-r_-^2)}{r_0^3}.
\end{equation}

We solve the Klein-Gordon equation
\begin{equation}
[(\nabla^\mu-iqA^\mu)(\nabla_\mu-iqA_\mu)-m^2]\Phi=0
\label{eq:KGeqn}
\end{equation}
for the charged scalar field $\Phi$ in the above background to obtain the Green's function of the dual scalar operator in the CFT. After the separation of variables
\begin{equation}
\Phi(r,x^\mu)\sim e^{-i\omega t}Y(\sigma)\phi(r)\,,
\label{eq:sep-phi}
\end{equation}
where $Y(\sigma)$ satisfies $\hat{\nabla}^2Y=-\bar{\lambda}Y$ with $\hat{\nabla}^2$ being the Laplacian on $d\Sigma_{3,\kappa}^2$, we obtain an ODE for $\phi(r)$. The function $Y(\sigma)$ depends on the value of $\kappa$: (i) for $\kappa=1$, $Y(\sigma)$ is the spherical harmonics on $\mathbb{S}^3$ with $\bar{\lambda}=l(l+2)$, where $l$ is the angular mode number; (ii) for $\kappa=0$, $Y(\sigma)$ is the plane waves with $\bar{\lambda}=k^2$, where $k$ is the wave number; (iii) for $\kappa=-1$, $Y(\sigma)$ is the hyperbolic harmonics on $\mathbb{H}^3$. For a normalizable mode on $\mathbb{H}^3$, we need $\bar{\lambda}\geq 1$. We will study the spherical black hole in this paper.

The equation of motion for $\phi$ is
\begin{equation}
\phi''+\left(\frac{f'}{f}+\frac{3}{r}\right)\phi'+\left(\frac{(\omega+qA_t)^2}{f^2}-\frac{\bar{\lambda}}{r^2f}
-\frac{m^2}{f}\right)\phi=0\,.\label{eq:KG-RN}
\end{equation}
To write it in terms of a Heun equation, we make the change of variables
\begin{equation}
    z=\frac{r^2-r_+^2}{r^2-r_-^2},\qquad t=\frac{r_0^2-r_+^2}{r_0^2-r_-^2}\,.
\end{equation}
The horizon is at $z=0$ and the AdS boundary is at $z=1$. We write
\begin{equation}
    \phi=z^{-1/2}(1-z)^{1/2}(1-z/t)^{-1/2}\psi(z)\,,
\end{equation}
where $\psi(z)$ satisfies the Heun equation in the normal form, i.e., \eqref{eq:heunEqN} below with parameters
\begin{gather}
    \theta_0=\frac{i\omega}{4\pi T_+},\qquad \theta_1=\frac{1}{2}\sqrt{m^2+4},\qquad \theta_t=\frac{i(\omega+qA_t(r_0))}{4\pi T_0},\qquad \theta_\infty=\frac{i(\omega+qA_t(r_-))}{4\pi T_-},\nonumber\\
    \mathsf{w}^2=\frac{(\omega+q\mu)^2-\bar{\lambda}+r_+^2+r_-^2-(m^2+2)r_0^2}{4(r_+^2-r_-^2)}+\theta_0^2\,.
\end{gather}

We impose the infalling boundary condition at the horizon,
\begin{equation}
    \phi\sim (r-r_+)^{-\frac{i\omega}{4\pi T_+}}\sim z^{-\theta_0}\,,
\end{equation}
where we have dropped the outgoing branch of solutions with $z^{\theta_0}$. The asymptotic behavior of the scalar field near the AdS boundary is
\begin{align}
\phi &=\frac{A}{r^{\Delta_-}}(1+\cdots)+\frac{B}{r^{\Delta_+}}(1+\cdots)+\frac{\tilde{B}\ln r}{r^{\Delta_+}}(1+\cdots)\nonumber\\
&\sim A(1-z)^{1-\theta_1}+B(1-z)^{1+\theta_1}+\tilde{B}(1-z)^{1+\theta_1}\ln(1-z)\,,\label{eq:phi-bdy}
\end{align}
where $\tilde{B}\neq 0$ when $\Delta_+-\Delta_-=2n$, where $n=1$, $2$, $\cdots$. When $\Delta_+=\Delta_-$, $\phi=Az^2\ln z+Bz^2+\cdots$. The retarded Green's function is
\begin{equation}
G=\frac{B}{A},
\end{equation}
where we have ignored an unimportant prefactor. When logarithm appears, extra terms must be added to above Green's function, but they do not change the imaginary part of the Green's function as well as the poles.  When $-4\leq m^2\leq -3$, there is an alternative quantization, by which the Green's function is $G=A/B$ \cite{Klebanov:1999tb}.

\subsection{Two-charge black hole in AdS$_5$ (Gubser-Rocha model)\label{sec:GR}}

The second system we study is the two-charge black hole AdS$_5$, which is a solution to the Gubser-Rocha model~\cite{Gubser:2009qt}. While the RN-AdS$_5$ black hole has finite entropy at zero temperature, the two-charge black hole in AdS$_5$ has zero entropy at zero temperature. In five-dimensional maximal gauged supergravity as a consistent truncation of type IIB supergravity, two charges of the three U(1) subgroups of the SO(6) gauge group are nonzero and equal, and the third is zero \cite{DeWolfe:2012uv,DeWolfe:2013uba}.

The action for the two-charge black hole in AdS$_5$ is
\begin{equation}
S=\int d^5x\sqrt{-g}\left(R-\frac{1}{4}e^{\frac{2}{\sqrt{6}}\bar{\phi}}F_{\mu\nu}^2-\frac{1}{2}(\partial_\mu\bar{\phi})^2+\frac{1}{L^2}\bigl(8e^{\frac{1}{\sqrt{6}}\bar{\phi}}+4e^{-\frac{2}{\sqrt{6}}\bar{\phi}}\bigr)\right),
\end{equation}
where the dilaton field $\bar{\phi}$ is a background neutral scalar field different from the perturbing charged scalar field $\Phi$. The metric for the black hole solution is
\begin{gather}
ds^2=e^{2\mathcal{A}}(-hdt^2+d\Sigma_{3,\kappa}^2)+\frac{e^{2\mathcal{B}}}{h}dr^2,\\
\mathcal{A}=\ln r+\frac{1}{3}\ln\biggl(1+\frac{Q^2}{r^2}\biggr),\qquad \mathcal{B}=-\ln r-\frac{2}{3}\ln\biggl(1+\frac{Q^2}{r^2}\biggr),\\
h=1+\frac{\kappa}{r^2+Q^2}-\frac{(r_+^2+Q^2)(r_+^2+Q^2+\kappa)}{(r^2+Q^2)^2}=\frac{(r^2-r_+^2)(r^2-r_0^2)}{(r^2+Q^2)^2},
\end{gather}
and the gauge field $A=A_tdt$ and the dilaton are
\begin{gather}
A_t=\mu\biggl(1-\frac{r_+^2+Q^2}{r^2+Q^2}\biggr),\qquad \mu=\sqrt{\frac{2(r_+^2+Q^2+\kappa)}{r_+^2+Q^2}}\,Q\,,\\
\bar{\phi}=\frac{2}{\sqrt{6}}\ln\biggl(1+\frac{Q^2}{r^2}\biggr),
\end{gather}
where $r_+$ is the horizon radius, $r_0^2=-\kappa-r_+^2-2Q^2$, and $Q$ is a parameter related to the chemical potential $\mu$. The parameter space of the solution is $(r_+, Q)$ for a specific $\kappa$.

The temperature of this black hole is
\begin{equation}
T_+=\left.\frac{|h'|e^{\mathcal{A}-\mathcal{B}}}{4\pi}\right|_{r=r_+}=\frac{r_+}{2\pi}\,\frac{r_+^2-r_0^2}{r_+^2+Q^2}=\frac{r_+}{2\pi}\biggl(2+\frac{\kappa}{r_+^2+Q^2}\biggr).
\end{equation}
In contrast to the RN-AdS$_5$ black hole, there is no inner horizon for spherical and planar cases, and the zero temperature limit is approached by $r_+=0$. Since $r=0$ is a spacetime singularity, the zero temperature limit has a naked singularity, which will be cloaked by a horizon at finite temperature. A way to resolve this singularity is by a ten-dimensional lift; see~\cite{Gubser:2009qt,Cvetic:1999xp}. The ``temperature'' at $r_0$ is
\begin{equation}
    T_0=\frac{r_0}{2\pi}\,\frac{r_0^2-r_+^2}{r_0^2+Q^2}.
\end{equation}

To obtain the Green's function for a scalar operator in the dual CFT, we solve the Klein-Gordon equation \eqref{eq:KGeqn} for a charged scalar field $\Phi$. After the separation of variables \eqref{eq:sep-phi}, the equation of motion for $\phi$ is
\begin{equation}
\phi''+\left(4\mathcal{A}'-\mathcal{B}'+\frac{h'}{h}\right)\phi'+\left(\frac{(\omega+qA_t)^2}{e^{2(\mathcal{A}-\mathcal{B})}h^2}-\frac{\bar{\lambda}}{e^{2(\mathcal{A}-\mathcal{B})}h}-\frac{m^2}{e^{-2\mathcal{B}}h}\right)\phi=0\,.
\label{eq:KG-GR}
\end{equation}
To write \eqref{eq:KG-GR} with $m=0$ in terms of a Heun equation, we make the change of variables
\begin{equation}
    z=1-\frac{r_+^2}{r^2},\qquad t=1-\frac{r_+^2}{r_0^2}\,.
\end{equation}
The horizon is at $z=0$ and the AdS boundary is at $z=1$. We write $\phi=z^{-1/2}(1-z)^{1/2}(1-z/t)^{-1/2}\psi(z)$, where $\psi(z)$ satisfies the Heun equation in the normal form, i.e., \eqref{eq:heunEqN} below with parameters
\begin{gather}
    \theta_0=\frac{i\omega}{4\pi T_+},\qquad \theta_1=1,\qquad \theta_t=\frac{i(\omega+qA_t(r_0))}{4\pi T_0},\qquad \theta_\infty=0,\nonumber\\
    \mathsf{w}^2=\frac{(\omega+q\mu)^2-\bar{\lambda}+r_+^2-2r_0^2}{4r_+^2}+\theta_0^2\,.
\end{gather}

After imposing the infalling condition at the horizon, the solution is $\psi^{[0]}_{+}$ in \eqref{eq:psi0pm-sol} below (with logarithmic terms), and the near boundary behavior is in \eqref{eq:psi1pm-sol} below.

As a side note, the above two equations \eqref{eq:KG-RN} and \eqref{eq:KG-GR} are solvable in terms of hypergeometric equations at zero temperature and $\omega=0$, leading to analytic solutions of the zero modes. For the extremal RN-AdS$_5$ black hole, the $\omega=0$ solution was obtained in \cite{Ren:2012hg}, and the Gubser-Rocha model, the $\omega=0$ solution was obtained in \cite{Alishahiha:2012ad}.  Those results are generalized to hyperbolic or spherical cases in \cite{Ren:2020djc}. The next order of finite $T$ and $\omega$ corrections for the planar RN-AdS$_5$ black hole was analytically obtained in \cite{Arnaudo:2024sen}.

\section{Heun equation and its connection formula from recursions}
\label{sec:Heun}

\subsection{Heun equation and connection formula}
The standard form of the Heun equation is
\begin{equation}
    \frac{d^{2}y}{dz^{2}}+\left(\frac{\gamma}{z}+\frac{\delta}{z-1}+\frac{\epsilon}{z-t}\right)\frac{dy}{dz}+\frac{\alpha\beta z-\mathsf{q}}{z(z-1)(z-t)}y=0\,,
\end{equation}
where $\epsilon=\alpha+\beta-\gamma-\delta+1$ and $\mathsf{q}$ is the accessory parameter. The four regular singular points $\{0,1,t,\infty\}$ and the corresponding exponents are represented by the Riemann scheme as
\begin{equation}
    y=P
    \setlength\arraycolsep{5pt}
    \begin{Bmatrix}
    0 & 1 & t & \infty &\\
    0 & 0 & 0 & \alpha & ;z\\
    1-\gamma & 1-\delta & 1-\epsilon & \beta &
    \end{Bmatrix}.
\end{equation}
The two linearly independent solutions near $z=0$ are
\begin{align}
    y^{[0]}_{1} &=H\!\ell (t,\mathsf{q};\alpha,\beta,\gamma,\delta;z),\\
    y^{[0]}_{2} &=z^{1-\gamma}\mathit{H\!\ell}\left(t,(t\delta+\epsilon)(1-\gamma)+\mathsf{q};\alpha+1-%
\gamma,\beta+1-\gamma,2-\gamma,\delta;z\right),
\end{align}
where $H\!\ell$ denotes a local Heun function, i.e., it satisfies one boundary condition at $z=0$, but generally does not satisfy the boundary condition at $z=1$. The two linearly independent solutions near $z=1$ are
\begin{align}
    y^{[1]}_{1} &=H\!\ell (1-t,\alpha\beta-\mathsf{q}; \alpha, \beta, \delta, \gamma; 1-z),\\
    y^{[1]}_{2} &=(1-z)^{1-\delta}\*\mathit{H\!\ell}\bigl(1-t,((1-t)\gamma+\epsilon)(1-\delta)+
\alpha\beta-\mathsf{q};\nonumber\\
&\hspace{0.4\textwidth}\alpha+1-\delta,\beta+1-\delta,2-\delta,\gamma;1-z\bigr).
\end{align}
Solutions to boundary value problems between $z=0$ and $z=1$ involve the connection formula between the two sets of solutions.

The normal form of the Heun equation is
\begin{multline}
   \Biggl(\frac{d^2}{dz^2}+\frac{\frac14-\theta_0^2}{z^2}+\frac{\frac14-\theta_1^2}{(z-1)^2}+\frac{\frac14-\theta_t^2}{(z-t)^2}+\frac{\theta_0^2+\theta_1^2+\theta_t^2-\theta_\infty^2-\tfrac12}{z(z-1)}\\
    +\frac{(t-1)(\mathsf{w}^2+\theta_t^2-\theta_\infty^2-\tfrac14)}{z(z-1)(z-t)}\Biggr)\psi(z)=0\,,
\label{eq:heunEqN}
\end{multline}
which is related to the standard form by $\psi(z)=z^{1/2-\theta_0}(1-z)^{1/2-\theta_1}(1-z/t)^{1/2-\theta_t}y(z)$ and
\begin{gather}
    \theta_0=\frac{1}{2}(1-\gamma),\qquad \theta_1=\frac{1}{2}(1-\delta),\qquad \theta_t=\frac{1}{2}(1-\epsilon),\qquad \theta_\infty=\frac{1}{2}(\beta-\alpha),\nonumber\\
    (t-1)\mathsf{w}^2+\mathsf{q}+\theta_0^2+\theta_\infty^2-t\bigl(\theta_0+\theta_1-\tfrac{1}{2}\bigr)^2-\bigl(\theta_0+\theta_t-\tfrac{1}{2}\bigr)^2=0\,.
\end{gather}
The four regular singular points and the corresponding exponents are represented by the Riemann scheme as
\begin{equation}
    \psi=P
    \setlength\arraycolsep{5pt}
    \begin{Bmatrix}
    0 & 1 & t & \infty &\\
    \frac{1}{2}-\theta_0 & \frac{1}{2}-\theta_1 & \frac{1}{2}-\theta_t & \frac{1}{2}-\theta_\infty & ;z\\
    \frac{1}{2}+\theta_0 & \frac{1}{2}+\theta_1 & \frac{1}{2}+\theta_t & \frac{1}{2}+\theta_\infty &
    \end{Bmatrix}.
\end{equation}
The two linearly independent solutions near $z=0$ are
\begin{align}
    \psi^{[0]}_\pm= &z^{\frac{1}{2}\mp\theta_0}(1-z)^{\frac{1}{2}-\theta_1}(1-z/t)^{\frac{1}{2}-\theta_t} H\!\ell\bigl(t,1/4+(1-t)\mathsf{w}^2+t(1/2\mp\theta_0-\theta_1)^2+\theta_t^2-\theta_\infty^2\nonumber\\
    &\mp\theta_0(1-2\theta_t)-\theta_t;1\mp\theta_0-\theta_1-\theta_t-\theta_\infty,
    1\mp\theta_0-\theta_1-\theta_t+\theta_\infty,1\mp 2\theta_0,1-2\theta_1,z\bigr).
\label{eq:psi0pm-sol}
\end{align}
The two linearly independent solutions near $z=1$ are
\begin{small}
\begin{align}
    &\psi^{[1]}_\pm= (1-z)^{\frac{1}{2}\mp\theta_1}z^{\frac{1}{2}-\theta_0}\Bigl(\frac{t-z}{t-1}\Bigr)^{\frac{1}{2}-\theta_t} H\!\ell\bigl(1-t,3/4-(1-t)\mathsf{w}^2-t(1/2-\theta_0\mp\theta_1)^2+\theta_0^2+\theta_1^2\nonumber\\
    &\mp 2\theta_1(1-\theta_0-\theta_t)-\theta_0-\theta_t;1-\theta_0\mp\theta_1-\theta_t-\theta_\infty,1-\theta_0\mp\theta_1-\theta_t+\theta_\infty,1\mp 2\theta_1, 1-2\theta_0;1-z\bigr).
\label{eq:psi1pm-sol}
\end{align}
\end{small}

We assume that $|t|>1$ and $\theta_0,\theta_1\notin\mathbb Z/2$. Let $\psi^{[0]}_{\pm}(z)$, $\psi^{[1]}_{\pm}(z)$ denote its Frobenius solutions normalized as in \eqref{eq:psi0pm} and \eqref{eq:psi1pm}. The relation between the two bases is given by \eqref{eq:psiconn}, where the function $\mathsf C(\theta_0,\theta_1)$ can be written as \cite{Lisovyy:2022flm}
\begin{equation}
    \mathsf{C}( \theta _{0} ,\theta _{1}) =\frac{\Gamma ( 1-2\theta _{0}) \Gamma ( 2\theta _{1})}{\Gamma \left(\frac{1}{2} +\theta _{1} -\theta _{0} +\mathsf{w}\right) \Gamma \left(\frac{1}{2} +\theta _{1} -\theta _{0} -\mathsf{w}\right)} (1-\lambda )^{\frac{1}{2} -\theta _{t}} a_{\infty },
\end{equation}
where $\lambda\equiv 1/t$ and $a_\infty=\lim_{k\to\infty}a_{k}$ of the following recurrence relation
\begin{equation}
a_{k+1} -a_{k} =-\lambda ( \alpha _{k} a_{k} +\beta _{k} a_{k-1}),\qquad a_{-1}=0,\qquad a_0=1, \label{eq:recur}
\end{equation}
where
\begin{subequations}\label{eq:alphabetaHE}
    	\begin{align}
    	\label{alphaHE}
    	\alpha_k=&\, -\frac{(k+\frac12-\theta_0-\theta_t)^2-\theta_0^2-\theta_\infty^2+\mathsf{w}^2}{( k+\frac12-\theta_0+\theta_1)^2-\mathsf{w}^2},\\
    	\label{betaHE}
    	\beta_k=&\,\frac{k(k-2\theta_0)\bigl(
    		(k-\theta_0+\theta_1-\theta_t)^2-\theta_\infty^2\bigr)}{\bigl((k+\frac12-\theta_0+\theta_1)^2-\mathsf{w}^2\bigr)
    		\bigl((k-\frac12-\theta_0+\theta_1)^2-\mathsf{w}^2\bigr)}.
    	\end{align} 	
\end{subequations}

\subsection{Perturbative solution of the connection formula\label{sec:confps}}
This recurrence relation \eqref{eq:recur} can be solved as \cite{Lisovyy:2022flm}
\begin{align}
\ln a_{\infty } & =-\ln (1-\lambda )+\sum _{k=1}^{\infty }\ln\left( 1-\lambda \alpha _{k-1} -\frac{\lambda \beta _{k}}{1-\lambda \alpha _{k} -\frac{\lambda \beta _{k+1}}{1-\dotsc .}}\right)\nonumber\\
&\equiv -\ln (1-\lambda )+\ln a_{\infty }^{\prime },
\end{align}
where we introduce the notation $a_{\infty }^{\prime }$ for simplicity. We will demonstrate that $a_{\infty }^{\prime }$ can be expressed perturbatively in the following series:
\begin{align}
\ln a_{\infty }^{\prime } & =\ \ \ \ \ \varrho _{1} \lambda +\varrho _{2} \lambda ^{2} +\varrho _{3} \lambda ^{3} +\cdots ,\ \\
a_{\infty }^{\prime } & =1+ \rho _{1} \lambda +\rho _{2} \lambda ^{2} +\rho _{3} \lambda ^{3} +\cdots ,\ 
\end{align}
where the coefficients admit analytic expressions. We assume that the continued fraction can be expanded as follows:
\begin{align}
1-\lambda \alpha _{k-1} -\frac{\lambda \beta _{k}}{1-\lambda \alpha _{k} -\frac{\lambda \beta _{k+1}}{1-\dotsc }} & =1-\xi _{1}[ k] \lambda -\xi _{2}[ k] \lambda ^{2} -\cdots \label{eq:ExpanContiFrack}\\
1-\lambda \alpha _{k} -\frac{\lambda \beta _{k+1}}{1-\dotsc } & =1-\xi _{1}[ k+1] \lambda -\xi _{2}[ k+1] \lambda ^{2} -\cdots .\label{eq:ExpanContiFrackp1}
\end{align}
Substituting \eqref{eq:ExpanContiFrack} into \eqref{eq:ExpanContiFrackp1}, we derive the recurrence relation for the expansion coefficients:
\begin{align}
\xi _{1}[ k] & =\alpha_{k-1} +\beta_{k}\\
\xi _{2}[ k] & =\beta_{k}\xi _{1}[ k+1] \nonumber\\ 
\xi _{3}[ k] & =\beta_{k}\left( \xi _{1}[ k+1]^{2} +\xi _{2}[ k+1]\right) \nonumber\\ 
 & \ \vdots \nonumber\\ 
\xi _{n+1}[ k] & =\beta_{k}\Biggl(\sum _{|M|=n} C_{M} \xi _{m_{1}}[ k+1] \cdots \xi _{m_{j}}[ k+1]\Biggr) ,\label{eq:XiNClosed}
\end{align}
where $M=\{m_{1} ,\cdots ,m_{j}\}$ denotes a partition of $n$. The summation is over all partitions of $n$, with the coefficient $C_{M}$ given by
\begin{equation}
C_{M} =\frac{j!}{r_{1} !\cdots r_{i} !} ,
\end{equation}
where $j$ is the length of the partition $M$, and $r$ is the repetition number of an element in $M$. 
It is worth noting that the expression for $\xi_{1}[k]$ can be written as
\begin{equation}
\xi_{1}[k] = \frac{\#k^2+\#k+\#}{\left(\left( k+\frac{1}{2} -\theta _{0} +\theta _{1}\right)^{2} -\mathsf{w}^{2}\right)\left(\left( k-\frac{1}{2} -\theta _{0} +\theta _{1}\right)^{2} -\mathsf{w}^{2}\right)},
\end{equation}
where $\#$ denotes constants independent of $k$. Thus, the asymptotic behavior is given by
\begin{equation}
\lim _{k\rightarrow \infty }( k\xi _{1}[ k]) =0. \label{eq:asympXi1}
\end{equation}
We expand $\ln a_{\infty }^{\prime }$ as
\begin{equation}
\begin{aligned}
\ln a_{\infty }^{\prime } = & \sum _{k=1}^{\infty }\ln\left( 1-\xi _{1}[ k] \lambda -\xi _{2}[ k] \lambda ^{2} -\xi _{3}[ k] \lambda ^{3} \cdots \right)\\
= & -\left(\sum _{k=1}^{\infty } \xi _{1}[ k] \lambda +\sum _{k=1}^{\infty }\frac{1}{2}\left( \xi _{1}[ k]^{2} +2\xi _{2}[ k]\right) \lambda ^{2} +\cdots \right) ,
\end{aligned}
\end{equation}
which provides the following expressions for the coefficients:
\begin{equation}
\begin{cases}
\varrho _{1} & =-\sum _{k=1}^{\infty } \xi _{1}[ k]\\
\varrho _{2} & =-\sum _{k=1}^{\infty }\frac{1}{2}\left( \xi _{1}[ k]^{2} +2\xi _{2}[ k]\right)\\
 & \ \vdots 
\end{cases} \quad\Longrightarrow\quad \begin{cases}
\rho _{1} & =\varrho _{1}\\
\rho _{2} & =\frac{1}{2}\left( \varrho _{1}^{2} +2\varrho _{2}\right)\\
 & \ \vdots 
\end{cases}.
\end{equation}
From \eqref{eq:XiNClosed}, we know that $\xi_{n+1}[k]$ can be expressed as $\beta_k$ multiplied by a sum of terms involving $\xi _{m_{1}}[ k+1] \cdots \xi _{m_{j}}[ k+1]$, where $m_{1} ,\cdots m_{j}$ are smaller than $n+1$. Therefore, by combining \eqref{eq:asympXi1} and $\lim_{k\rightarrow \infty }\beta_k=1$, we find that
\begin{equation}
\lim _{k\rightarrow \infty }( k\xi _{n+1}[ k]) =0\,, \label{eq:XiNLimit}
\end{equation}
which ensures that the summations in $\varrho _{n+1}$, expressed by $\displaystyle \xi _{n+1}[ k]$, are smaller than a harmonic series. These summations can be analytically expressed using the polygamma function after performing partial fraction decomposition:
\begin{equation}
\varrho _{n} =\sum _{k=1}^{\infty }\left(\sum _{i=1}^{p}\frac{a_{i}}{( k+b_{i})^{r_{i}}}\right) =\sum _{i=1}^{p}\frac{(-1)^{r_{i}}}{( r_{i} -1) !} a_{i} \psi ^{( r_{i} -1)}( b_{i} +1) .\ 
\end{equation}
where $\psi ^{( r_{i} -1)}$ is the polygamma function of order $r_{i} -1$. Following this procedure, we obtain:
\begin{equation}
\begin{aligned}
\varrho _{1} = & \frac{1}{16\mathsf{w}\left( -1+4\mathsf{w}^{2}\right)}\bigl[( 1-4\mathsf{w}^{2} +4\theta _{0}^{2} -4\theta _{1}^{2})( -1+4\mathsf{w}^{2} +4\theta _{t}^{2} -4\theta _{\infty }^{2}) \psi ^{( 0)}\Bigl(\frac{1}{2} -\mathsf{w} -\theta _{0} +\theta _{1}\Bigr)\\
 & +\left( 1-4\mathsf{w}^{2} +4\theta _{0}^{2} -4\theta _{1}^{2}\right)\left( 1-4\mathsf{w}^{2} -4\theta _{t}^{2} +4\theta _{\infty }^{2}\right) \psi ^{( 0)}\left(\frac{1}{2} +\mathsf{w} -\theta _{0} +\theta _{1}\right)\\
 & +8\mathsf{w}\left( 1+\theta _{0} +\theta _{1} +2\theta _{t} -4\theta _{0} \theta _{t}^{2} -4\theta _{1} \theta _{t}^{2} -4\mathsf{w}^{2}( 1+\theta _{0} +\theta _{1} +2\theta _{t}) +4\theta _{0} \theta _{\infty }^{2} +4\theta _{1} \theta _{\infty }^{2}\right)\bigr] .
\end{aligned}
\end{equation}

There are two folds of infinity in the recurrence relation: an infinite number of recursions and an infinite order of $\lambda$. (i) For general $\lambda$ (not necessarily small), we can truncate the recurrence relation by $n$. This is not a perturbative solution, but an approximation with a finite number of recurrences. (ii) For a specific order of $\lambda$, the infinite number of recursions can be analytically summed. This is the perturbative solution of the recurrence relation.

\subsection{Connection formula with logarithms}
In physics problems, it is common to have logarithmic terms in the Frobenius solutions, which happens when $\theta_0, \theta_1\notin\mathbb{Z}/2$ are not satisfied. For $\theta_0\notin\mathbb{Z}/2$ and $\theta_1\in\mathbb{Z}/2$, the two series \eqref{eq:psi1pm} are no longer linearly independent, and we must change the basis. Start with the connection formula for $\theta_1\notin\mathbb{Z}/2$, we insert an identity matrix $\mathsf{M}^{-1}\mathsf{M}$ to change the basis:\footnote{We thank Oleg Lisovyy for pointing out this elegant treatment. The connection formula with logarithmic terms was also addressed by \cite{He:2023wcs,Jia:2024zes} in slightly different approaches.}
\begin{equation}
\setlength\arraycolsep{5pt}
    \begin{pmatrix}
         \psi_1^{[0]}(z)\\
         \psi_2^{[0]}(z) 
    \end{pmatrix}=
    \begin{pmatrix}
         \mathsf{C}_{11} & \mathsf{C}_{12}\\
         \mathsf{C}_{21} & \mathsf{C}_{22}
    \end{pmatrix}\mathsf{M}^{-1}\mathsf{M}
    \begin{pmatrix}
         \psi_1^{[1]}(z)\\
         \psi_2^{[1]}(z) 
    \end{pmatrix}\equiv
    \tilde{\mathsf{C}}\begin{pmatrix}
         \tilde{\psi}_1^{[1]}(z)\\
         \psi_2^{[1]}(z) 
    \end{pmatrix},
\end{equation}
where the new basis and the corresponding connection matrix are
\begin{equation}
\setlength\arraycolsep{5pt}
    \begin{pmatrix}
         \tilde{\psi}_1^{[1]}(z)\\
         \psi_2^{[1]}(z) 
    \end{pmatrix}=
    \mathsf{M}\begin{pmatrix}
         \psi_1^{[1]}(z)\\
         \psi_2^{[1]}(z) 
    \end{pmatrix},\qquad
    \tilde{\mathsf{C}}=\begin{pmatrix}
         \mathsf{C}_{11} & \mathsf{C}_{12}\\
         \mathsf{C}_{21} & \mathsf{C}_{22}
    \end{pmatrix}\mathsf{M}^{-1}\,.
\end{equation}
The cases $\theta_1=0$ and $\theta_1=n/2$ $(n=1,2,\cdots)$ are treated differently as follows.
\begin{itemize}
    \item $\theta_1=0$. Starting with $\theta_1\neq 0$, we can choose $\mathsf{M}$ as
    \begin{equation}
\setlength\arraycolsep{5pt}
    \mathsf{M}=\begin{pmatrix}
         1/\theta_1 & -1/\theta_1\\
         0 & 1
    \end{pmatrix},\qquad
    \mathsf{M}^{-1}=\begin{pmatrix}
         \theta_1 & 1\\
         0 & 1
    \end{pmatrix}.
\end{equation}
Upon taking the $\theta_1\to 0$ limit, we obtain a new function $\tilde{\psi}^{[1]}_1$ different from $\psi^{[1]}_2$:
\begin{equation}
    \tilde{\psi}^{[1]}_1=\lim_{\theta_1\to 0}\frac{\psi^{[1]}_1-\psi^{[1]}_2}{\theta_1}=\frac{\partial\psi^{[1]}_1}{\partial\theta_1}\bigg\vert_{\theta_1=0}.
\end{equation}
The new connection matrix is given by
\begin{equation}
\setlength\arraycolsep{5pt}
    \tilde{\mathsf{C}}=\lim_{\theta_1\to 0}
    \begin{pmatrix}
         \mathsf{C}_{11} & \mathsf{C}_{12}\\
         \mathsf{C}_{21} & \mathsf{C}_{22}
    \end{pmatrix}\mathsf{M}^{-1}=\lim_{\theta_1\to 0}
    \begin{pmatrix}
         \theta_1\mathsf{C}_{11} & \mathsf{C}_{11}+\mathsf{C}_{12}\\
         \theta_1\mathsf{C}_{21} & \mathsf{C}_{21}+\mathsf{C}_{22}
    \end{pmatrix}.
\end{equation}

    \item $\theta_1=n/2$, where $n=1, 2, \cdots$.
Starting with $\theta_1=n/2+\delta\theta$, we choose $\mathsf{M}$ as
\begin{equation}
\setlength\arraycolsep{5pt}
    \mathsf{M}=\begin{pmatrix}
         1 & -\varphi^{[1]}_{1,n}\\
         0 & 1
    \end{pmatrix},\qquad
    \mathsf{M}^{-1}=\begin{pmatrix}
         1 & \varphi^{[1]}_{1,n}\\
         0 & 1
    \end{pmatrix}
\end{equation}
where $\varphi^{[1]}_{1,n}$ is the coefficient in
\begin{equation}
    \psi^{[1]}_1=(1-z)^{1/2-\theta_1}\bigl[1+\varphi^{[1]}_{1,1}(1-z)+\cdots+\varphi^{[1]}_{1,n}(1-z)^n+\cdots\bigr].
\end{equation}
The coefficients starting from $\varphi^{[1]}_{1,n}$ are singular at $\theta_1=n/2$. The above choice of $\mathsf{M}$ ensures that there is no $(1-z)^n$ term in $\tilde{\psi}^{[1]}_1$.
The function $\tilde{\psi}^{[1]}_1$ is given by the limit, 
\begin{equation}
    \tilde{\psi}^{[1]}_1=\lim_{\delta\theta\to 0}\Bigl(\psi^{[1]}_1-\varphi^{[1]}_{1,n}\psi^{[1]}_2\Bigr),
\end{equation}
which comprises two parts, one of which is $\psi^{[1]}_2$ multiplied by $\ln(1-z)$, and the other has the series expansion coefficients given by the limit $\delta\theta\to 0$:
\begin{equation}
\begin{aligned}
\tilde{\psi}^{[1]}_1 = &(1-z)^{1/2-n/2}\lim _{\delta\theta\to 0}\Bigl[1+\varphi_{1,1}^{[1]}(1-z)+\cdots+ \varphi_{1,n-1}^{[1]}(1-z)^{n-1}\\
&+\sum_{k=1}^{\infty}\left( \varphi _{1,n+k}^{[1]}-\varphi_{1,n}^{[1]} \varphi_{2,k}^{[1]}\right)(1-z)^{n+k}\Bigr]-2\ln(1-z)\lim_{\delta\theta\to 0}\Bigl(\delta\theta\,\varphi_{1,n}^{[1]}\Bigr)\psi_2^{[1]}.
\end{aligned}
\end{equation}
The new connection matrix is given by
\begin{equation}
\setlength\arraycolsep{5pt}
    \tilde{\mathsf{C}}=\lim_{\delta\theta\to 0}
    \begin{pmatrix}
         \mathsf{C}_{11} & \mathsf{C}_{12}\\
         \mathsf{C}_{21} & \mathsf{C}_{22}
    \end{pmatrix}\mathsf{M}^{-1}=\lim_{\delta\theta\to 0}
    \begin{pmatrix}
         \mathsf{C}_{11} & \mathsf{C}_{11}+\varphi^{[1]}_{1,n}\mathsf{C}_{12}\\
         \mathsf{C}_{21} & \mathsf{C}_{21}+\varphi^{[1]}_{1,n}\mathsf{C}_{22}
    \end{pmatrix},
\end{equation}
where $\mathsf{C}_{11}$ and $\mathsf{C}_{21}$ are regular at $\delta\theta=0$.
\end{itemize}

\subsection{Applications to thermal correlators}

\begin{figure}
  \centering
  \includegraphics[width=0.32\linewidth]{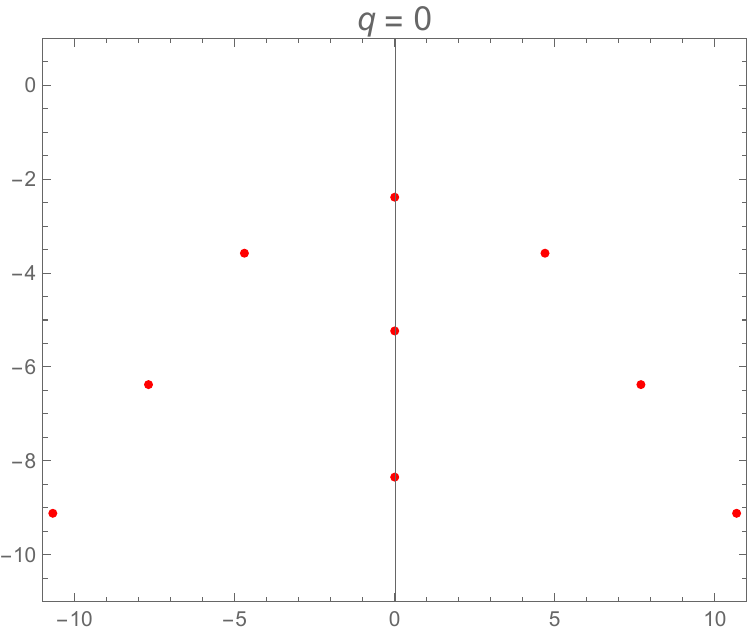}
  \includegraphics[width=0.32\linewidth]{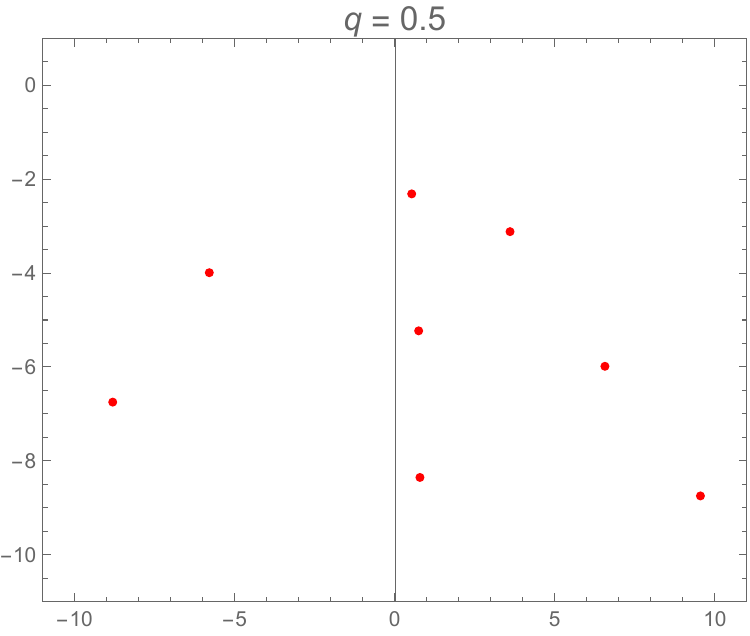}
  \includegraphics[width=0.32\linewidth]{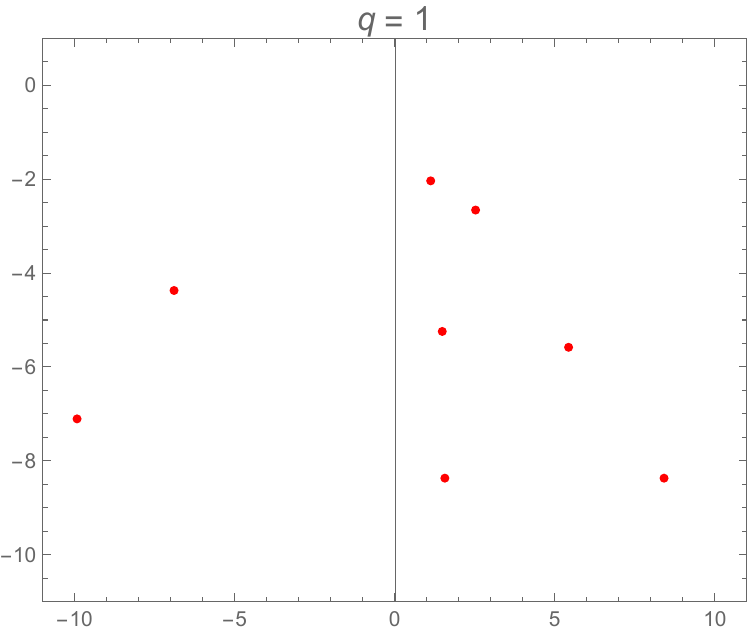}\\
  \vspace{3pt}
  \includegraphics[width=0.32\linewidth]{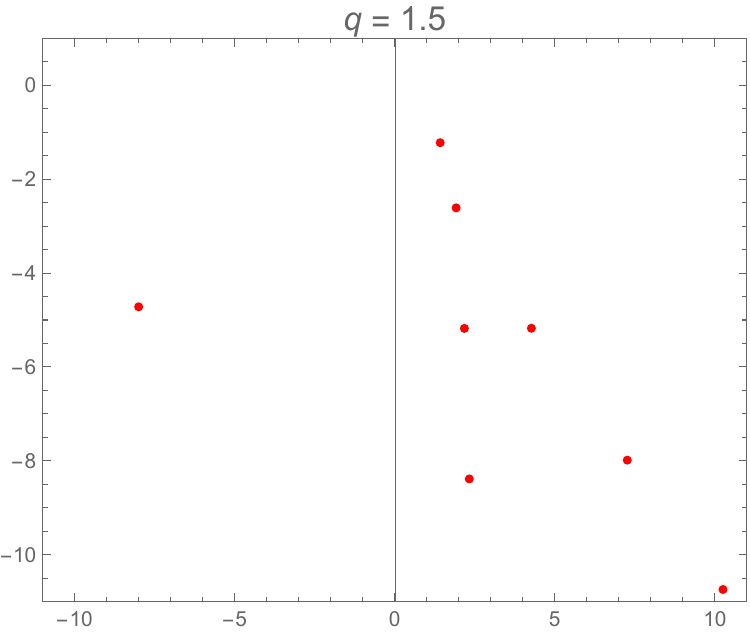}
  \includegraphics[width=0.32\linewidth]{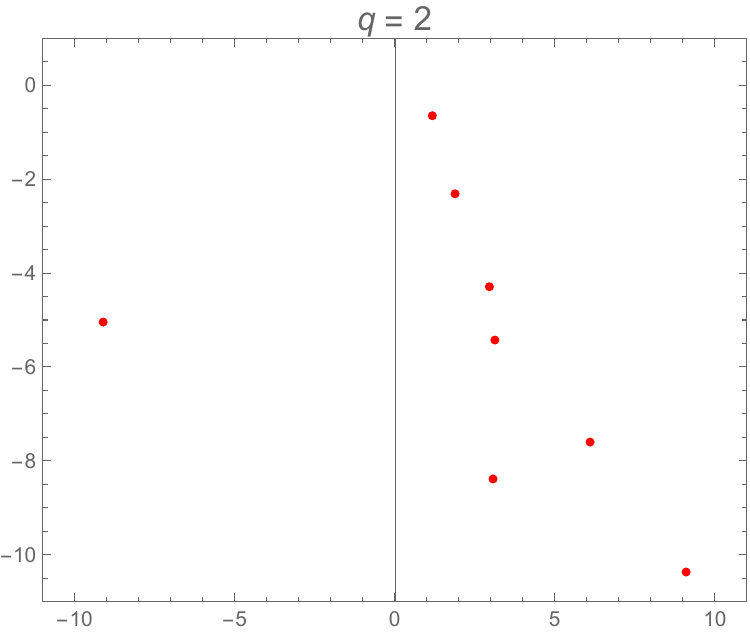}
  \includegraphics[width=0.32\linewidth]{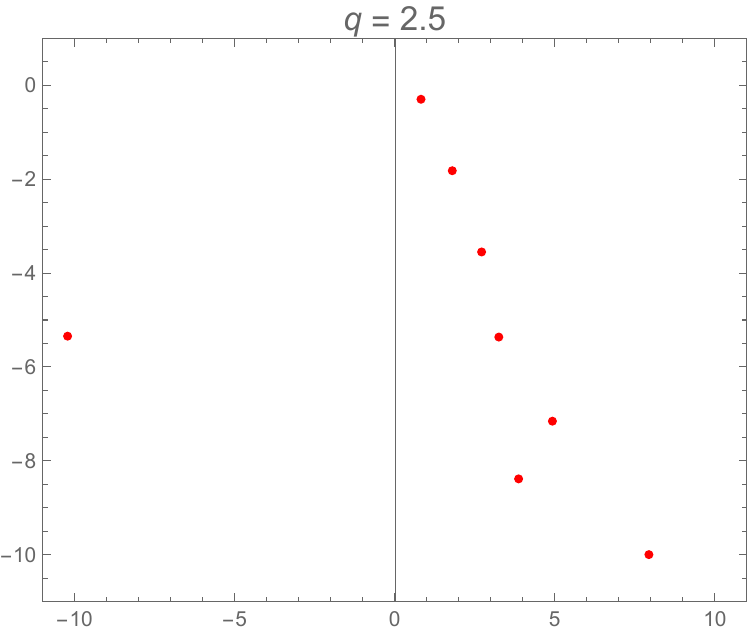}\\
  \vspace{3pt}
  \includegraphics[width=0.32\linewidth]{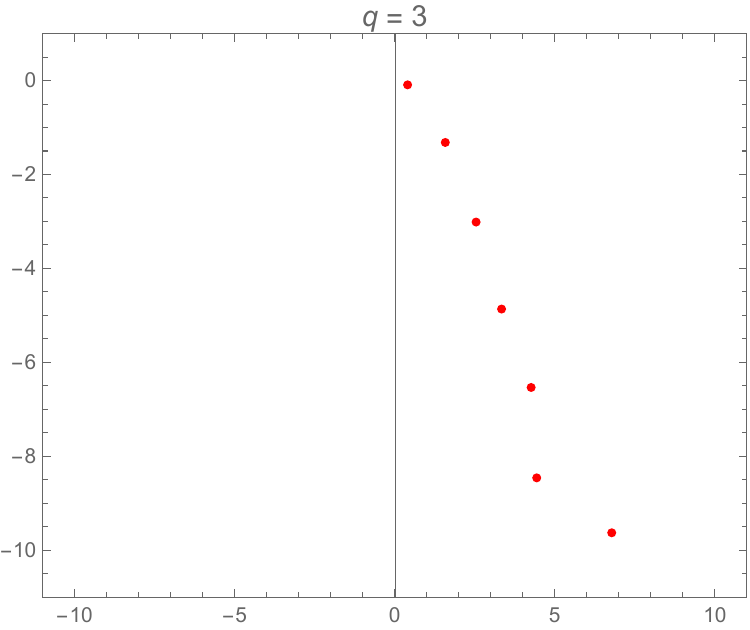}
  \includegraphics[width=0.32\linewidth]{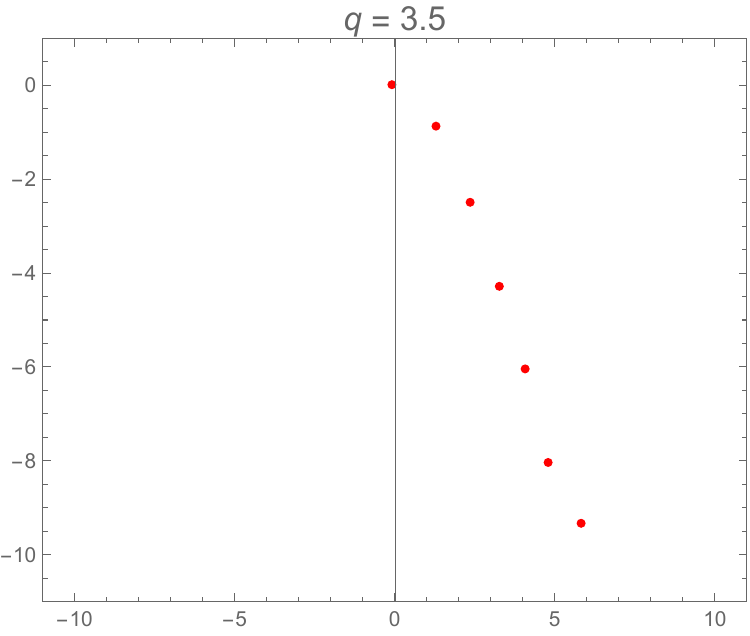}
  \includegraphics[width=0.32\linewidth]{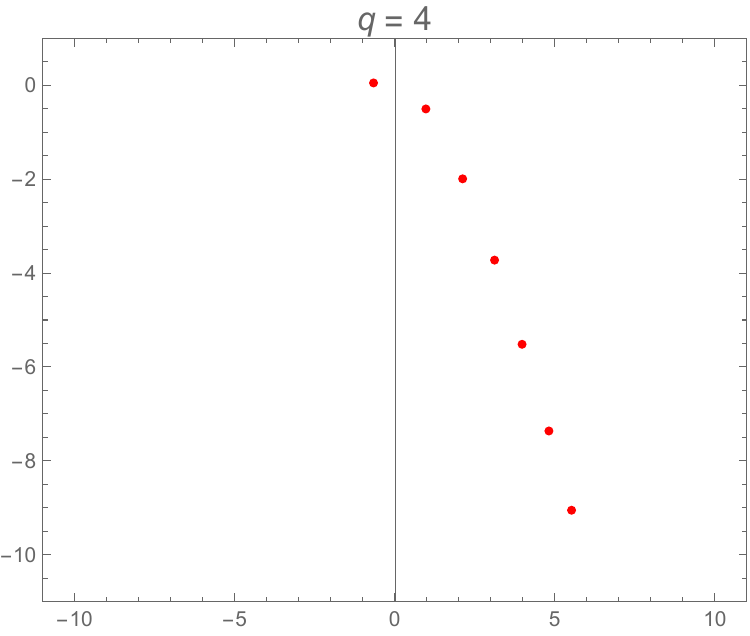}
  \caption{Quasinormal modes of the spherical RN-AdS$_5$ black hole perturbed by a charged scalar field. The horizontal axis is Re$(\omega)$ in the range $(-11,11)$, and the vertical axis is Im$(\omega)$ in the range $(-11,1)$. The parameters of the black hole are $(r_+,Q)=(1, 1.2)$, where the extremal limit is $Q_c=\sqrt{3}$. We take the angular mode number $l=0$ and the mass $m=0$ and increase the charge $q$ from zero. The plots show how the modes move as we increase the charge of the scalar field.}
  \label{fig:Q1}
\end{figure}

\begin{figure}
  \centering
  \includegraphics[width=0.32\linewidth]{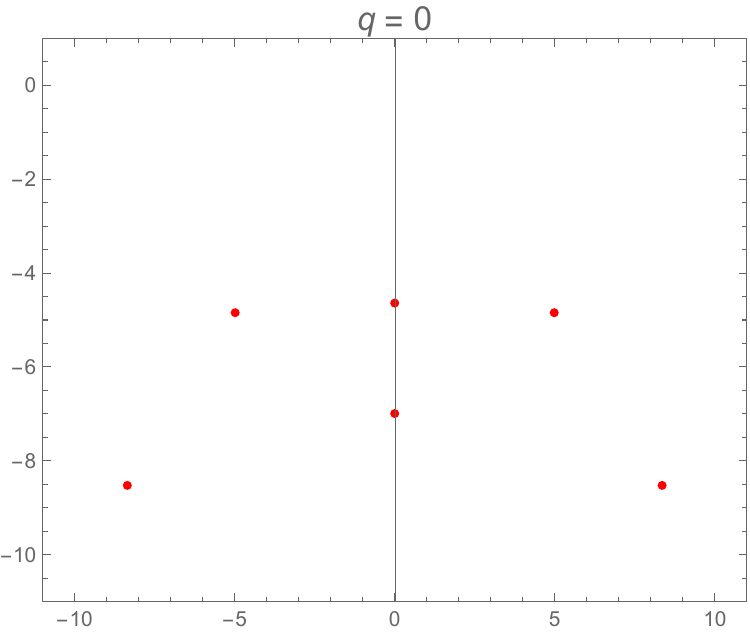}
  \includegraphics[width=0.32\linewidth]{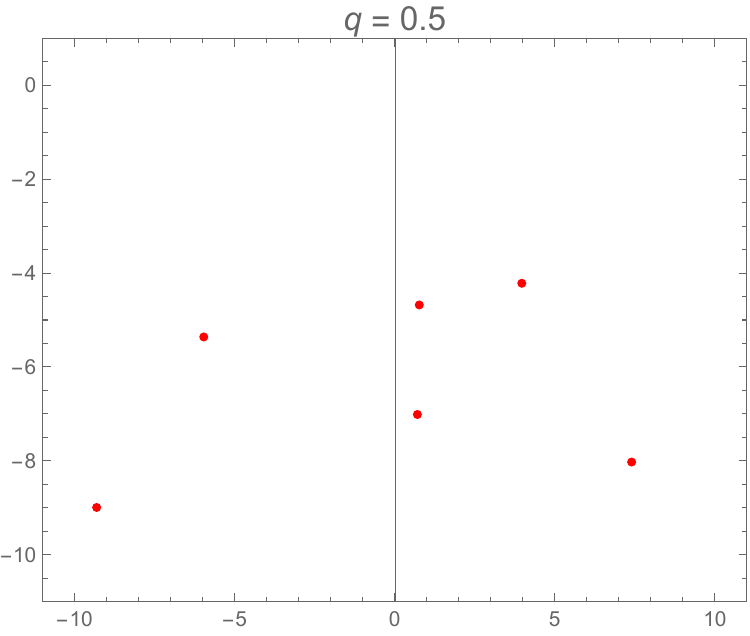}
  \includegraphics[width=0.32\linewidth]{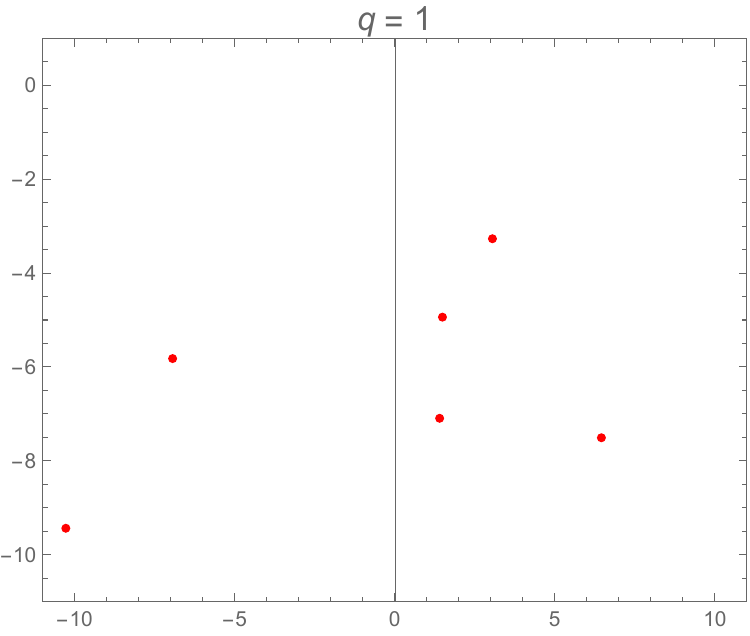}\\
  \vspace{3pt}
  \includegraphics[width=0.32\linewidth]{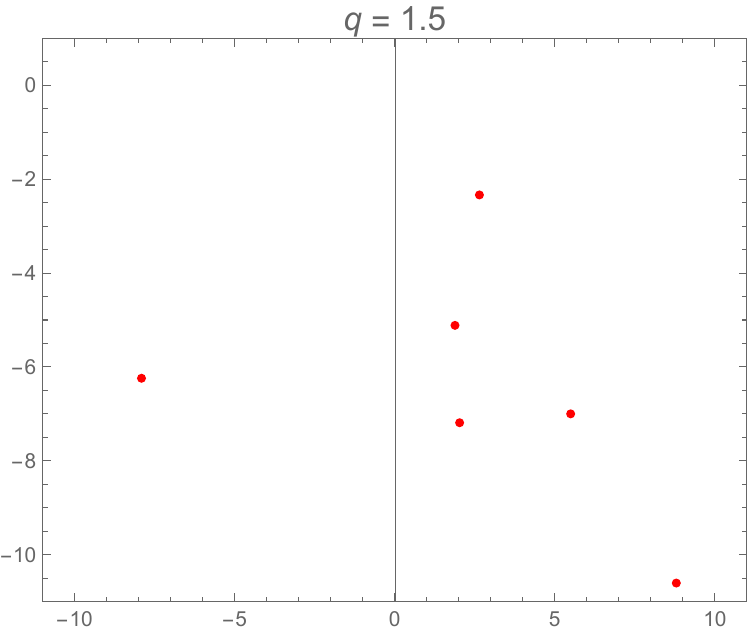}
  \includegraphics[width=0.32\linewidth]{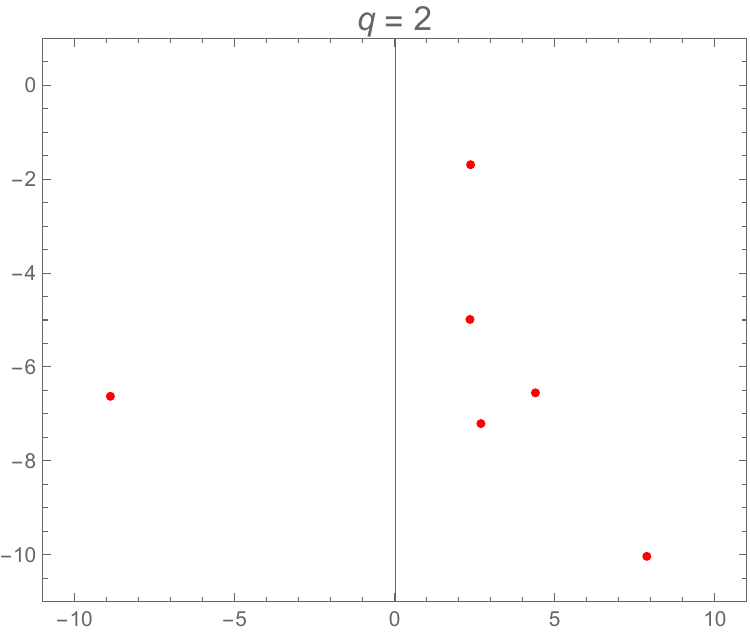}
  \includegraphics[width=0.32\linewidth]{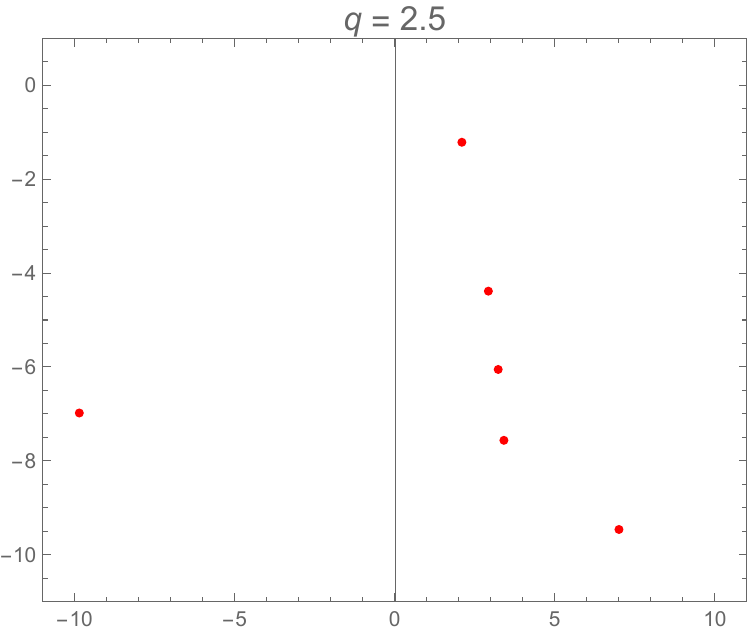}\\
  \vspace{3pt}
  \includegraphics[width=0.32\linewidth]{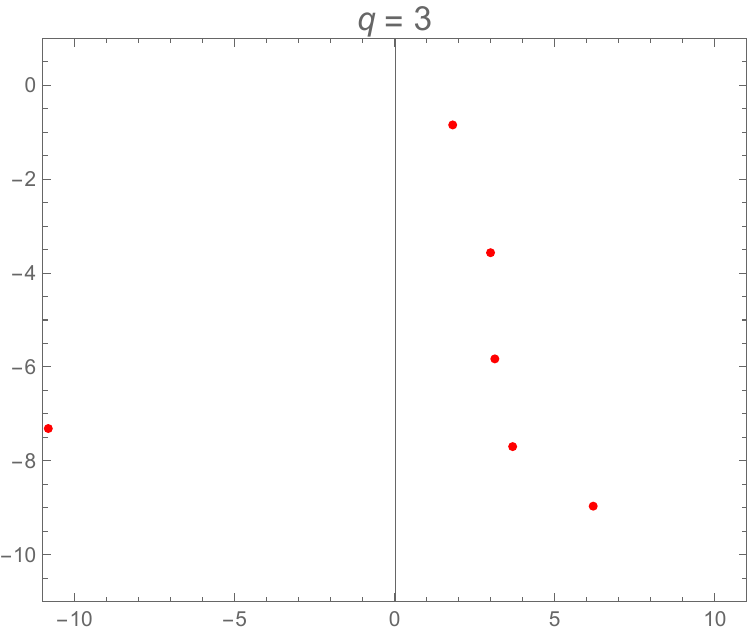}
  \includegraphics[width=0.32\linewidth]{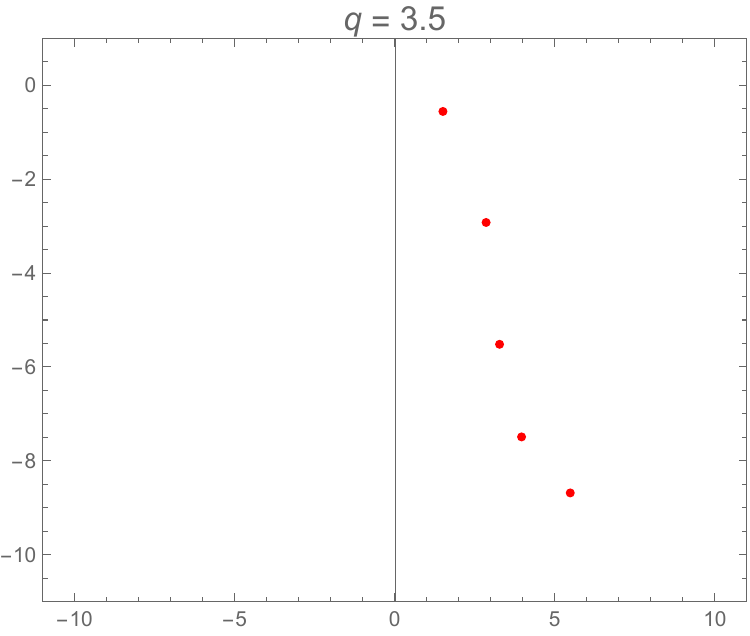}
  \includegraphics[width=0.32\linewidth]{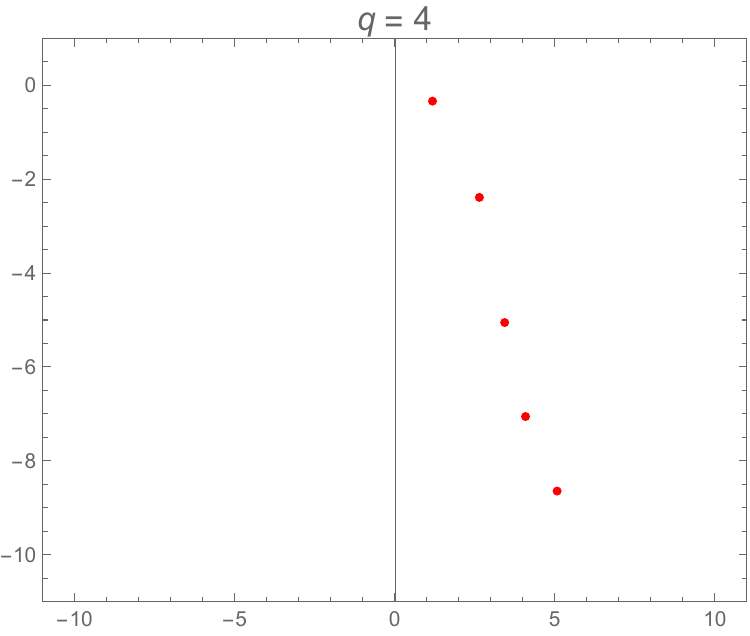}
  \caption{Quasinormal modes of the spherical two-charge black hole in AdS$_5$ perturbed by a charged scalar field. The horizontal axis is Re$(\omega)$ in the range $(-11,11)$, and the vertical axis is Im$(\omega)$ in the range $(-11,1)$. The parameters of the black hole are $(r_+,Q)=(0.8, 1.2)$. We take the angular mode number $l=0$ and the mass $m=0$ and increase the charge $q$ from zero. The plots show how the modes move as we increase the charge of the scalar field.}
  \label{fig:Q2}
\end{figure}

The parameters for the perturbation equation are $(\omega,\bar{\lambda},m^2,q)$, in addition to $(r_+,Q)$ for the black holes. Here we consider the spherical black hole with $\bar{\lambda}=l(l+2)$. While it is difficult to exhaust the full parameter space numerically, major features of the poles of the correlators, i.e., quasinormal modes of the black hole, are as follows.

\begin{itemize}
    \item For a charged black hole, there are two sets of quasinormal modes. One lies outside the imaginary axis, and the other lies on the imaginary axis. For neutral black holes, there are only poles outside the imaginary axis. These are known features of charged black holes \cite{Berti:2003ud,Janiszewski:2015ura}. Typical QNMs as $q$ is increased are shown in figure~\ref{fig:Q1} for the RN-AdS$_5$ black hole and figure~\ref{fig:Q2} for the 2-charge black hole in AdS$_5$.
    \item At fixed $r_+$, the poles in the imaginary become increasingly denser when we increase $Q$. A branch cut is expected in the extremal limit of the black hole, corresponding to the non-analyticity of the IR Green's function \cite{Faulkner:2009wj}.
    \item As we increase the charge $q$ of the scalar field, the poles become asymmetric. The poles originally on the imaginary axis merge with the poles outside the imaginary axis. There are eigenvalue repulsions, similar to the \cite{Dias:2022oqm,Davey:2023fin}.
    \item When $q$ is sufficiently large, a pole will move into the upper $\omega$-plane crossing the origin. This is the zero mode signaling instability. This is the instability of holographic superconductors \cite{Gubser:2008px,Hartnoll:2008vx,Hartnoll:2008kx} or superradiance \cite{Maeda:2010hf,Basu:2010uz}.
    \item For the two-charge black hole in AdS$_5$, the purely imaginary QNMs at $q=0$ are in pairs, which has been found in \cite{deOliveira:2024bgh}. If we decrease the temperature and increase $q$, more intricate behaviors will happen.
\end{itemize}

A tentative explanation is as follows. The two sets of poles at finite temperature are related to two types of instabilities at zero temperature. One is triggered by a zero mode discussed above, and the other is from the IR instability. In the IR Green's function, there is a branch cut along the negative imaginary $\omega$ axis, and the branch cut becomes poles at finite temperatures. For the RN-AdS$_5$ black holes, the poles on the imaginary axis reach the origin first at a sufficiently low temperature, which is consistent with the finding that the IR instability comes first at zero temperature. For the two-charge black hole in AdS$_5$, the IR scaling exponent is always real, and thus, the zero mode triggers the instability.

\section{Comparison to the Trieste formula}
\label{sec:Trieste}

The connection formula of Heun type equations expressed in terms of semiclassical Virasoro conformal blocks is obtained by \cite{Bonelli:2022ten,Bonelli:2021uvf}. In this section, we will briefly review the basic idea of the Trieste formula, give a derivation to it, and compare the recurrence relation with the Trieste formula. We take the notation used in \cite{Lisovyy:2022flm}, whose result is equivalent to \cite{Bonelli:2022ten,Bonelli:2021uvf}. Readers may skip this section without affecting the usage of the recurrence relation for the connection formula.

The key idea of the Trieste formula is to use the crossing symmetry of the conformal block to fuse the degenerate operator $\phi _{(2,1)}$ with two different operators and obtain the connection formula for two sets of solutions around singularities of BPZ equation. We consider a Liouville CFT on a sphere, whose central charge and conformal weight are parameterized by 
\begin{equation}
c=1+6(b+b^{-1} )^{2},\qquad h=\frac{1}{4} (b+b^{-1} )^{2} -p^{2} ,
\end{equation}
where $b$ is the coupling constant, and $p$ is the Liouville momentum. We are interested in an $n$-point correlation function expanded by conformal blocks:
\begin{equation}
\begin{aligned}
 & \langle \phi _{p_{n-1}}( z_{n-1}) \phi _{p_{n-2}}( z_{n-2}) \cdots \phi _{p_{1}}( z_{1}) \phi _{p_{0}}( z_{0}) \rangle \\
= & \sum _{[ s_{1}] ,\cdots [ s_{n-3}]}\bra{\phi _{p_{n-1}}} \phi _{p_{n-2}}\ket{\phi _{s_{n-3}}}\bra{\phi _{s_{n-3}}} \cdots \ket{\phi _{s_{1}}}\bra{\phi _{s_{1}}} \phi _{p_{1}}\ket{\phi _{p_{0}}}\\
= & \sum _{s_{1} ,\cdots s_{n-3}} C_{p_{n-1} p_{n-2}}^{s_{n-3}} C_{s_{n-3} p_{n-3}}^{s_{n-4}} \cdots C_{s_{2} p_{2}}^{s_{1}} C_{s_{1} p_{1}}^{p_{0}} \mathcal{F}( b;s_{1} ,\cdots ,s_{n-3} ;p_{0} ,\cdots p_{n-1} ; z_{2} ,\cdots z_{n-2}) ,
\end{aligned}
\end{equation}
where the sum over $[s]$ formally denotes the sum over all primary states $\ket{\phi _{s}}$ in the CFT spectrum as well as the corresponding orthogonalized and normalized descendants, and the antiholomorphic part has been ignored for convenience. The coordinates $z_{n-1}$, $z_{1}$, $z_{0}$ are fixed to $\infty$, $1$, $0$, respectively.

By inserting the null operator $\phi_{(2,1)}$ with conformal weight $h_{(2,1)} =-\frac{1}{2} -\frac{3}{4} b^{2}$ at $z$,\footnote{A null operator of type $\phi _{(r,s)}$ has momentum $\frac{1}{2}(rb+sb^{-1})$. See \cite{Ribault:2014hia} for a detailed review of CFTs in two dimensions.} we obtain the BPZ equation \cite{Belavin:1984vu}: 
\begin{equation}
\left(\frac{1}{b^{2}}\frac{\partial ^{2}}{\partial z^{2}} +\sum _{i=0}^{n-1}\left(\frac{1}{z-z_{i}}\frac{\partial }{\partial z_{i}} +\frac{h_{i}}{( z-z_{i})^{2}}\right)\right) \langle \phi _{( 2,1)}( z) \phi _{p_{n-1}}( z_{n-1}) \cdots \phi _{p_{1}}( z_{1}) \phi _{p_{0}}( z_{0}) \rangle =0.
\end{equation}
When $z_{n-1}$, $z_{1}$, and $z_{0}$ are fixed, the BPZ equation is simplified by replacing $\frac{\partial }{\partial z_{i}}$ for $i=0,1,n-1$ with appropriate terms. To achieve this, we utilize the following identity: 
\begin{equation}
\oint _{\infty }\frac{dy}{2\pi i}\frac{( y-z_{0})( y-z_{1})( y-z_{n-1})}{y-z} \langle T( y) \phi _{( 2,1)}( z) \phi _{p_{n-1}}( z_{n-1}) \cdots \phi _{p_{1}}( z_{1}) \phi _{p_{0}}( z_{0}) \rangle =0,
\end{equation}
in combination with the BPZ equation. The result is
\begin{equation}
\begin{aligned}
 & \biggl(\frac{1}{b^{2}}\frac{\partial ^{2}}{\partial z^{2}} -\left(\frac{1}{z} +\frac{1}{z-1}\right)\frac{\partial }{\partial z} +\sum _{k=2}^{n-2}\frac{z_{k}( z_{k} -1)}{z(z-1)( z-z_{k})}\frac{\partial }{\partial z_{k}} +\sum _{k=0}^{n-2}\frac{h_{k}}{( z-z_{k})^{2}}\\
 & \hspace{0.35\textwidth} +\frac{h_{n-1} -h_{(1,2)} -\sum _{k=0}^{n-2} h_{k}}{z(z-1)}\biggr) \langle \phi _{( 2,1)}( z) \phi \cdots \rangle =0.
\end{aligned}
\end{equation}
The degenerate operator can be absorbed into a regular conformal block, resulting in a degenerate block $\tilde{\mathcal{F}}$:
\begin{equation}
\begin{aligned}
 & \langle \phi _{( 2,1)}( z) \phi _{p_{n-1}}( z_{n-1}) \cdots \phi _{p_{1}}( z_{1}) \phi _{p_{0}}( z_{0}) \rangle \\
= & \sum _{s_{1} ,\cdots s_{n-3}} C_{p_{n-1} p_{n-2}}^{s_{n-3}} C_{s_{n-3} p_{n-3}}^{s_{n-4}} \cdots C_{s_{2} p_{2}}^{s_{1}} C_{s_{1} p_{1}}^{p_{0}}\tilde{\mathcal{F}}( b;s_{1} ,\cdots ,s_{n-3} ;p_{0} ,\cdots p_{n-1} ;z_{2} ,\cdots z_{n-2} ;z) ,
\end{aligned}
\end{equation}
where for $n=3$, a special case arises in which all partial derivatives with respect to $z_{k}$ vanish, reducing the BPZ equation to a hypergeometric equation. This hypergeometric equation is solvable due to the existence of connection formulas, which are also the fusion transformation formulas for the four-point degenerate conformal block.

The fusion rule\footnote{The fusion rule is a shorthand notation of operator product expansion (OPE), where the OPE coefficients are omitted. } for the degenerate operator $\phi _{(2,1)}$ with a non-degenerate operator $\phi _{p}$ is given by: 
\begin{equation}
\phi _{( 2,1)}\times\phi _{p} =\sum _{\epsilon =\pm } [\phi_{p+\epsilon b/2}].
\end{equation}
We denote the degenerate block obtained by fusing $\phi_{(2,1)}(z)$ with $\phi_{p_{0}}(z_{0})$ as $\tilde{\mathcal{F}}_{\epsilon}^{[0]}$, and the block obtained by fusing $\phi_{(2,1)}(z)$ with $\phi_{p_{1}} (z_{1} )$ as $\tilde{\mathcal{F}}_{\epsilon}^{[1]}$. The transformation formula between the four-point blocks $\tilde{\mathcal{F}}_{\epsilon}^{[0]}$ and $\tilde{\mathcal{F}}_{\epsilon}^{[1]}$ is known due to the connection formula of the hypergeometric equation:
\begin{equation}
\tilde{\mathcal{F}}_{\epsilon }^{[0]}( b;p_{0} ,p_{1} ,p_{\infty } ;z) =\sum _{\epsilon ^{\prime } =\pm } \mathsf{C}_{\text{hyp}}(\epsilon bp_{0} ,\epsilon^{\prime } bp_{1} ,bp_{\infty })\tilde{\mathcal{F}}_{\epsilon ^{\prime }}^{[1]}( b;p_{0} ,p_{1} ,p_{\infty } ;z) ,
\end{equation} where
\begin{equation}
\mathsf{C}_{\text{hyp}}(bp_{0} ,bp_{1} ,bp_{\infty })=\frac{\Gamma ( 1-2 bp_{0}) \Gamma \left( 2bp_{1}\right)}{\prod _{\pm } \Gamma \left(\frac{1}{2}-bp_{0} +bp_{1} \pm bp_{\infty}\right)} .\label{eq:hypercon1}
\end{equation}
When $n\geq 4$, a similar formula still holds, as the fusion transformation is a local behavior between $0$ and $1$: 
\begin{equation}
\begin{aligned}
 & \tilde{\mathcal{F}}_{\epsilon }^{[0]}( b;s_{1} ,\cdots ,s_{n-3} ;p_{0} ,\cdots p_{n-1} ;z_{2} ,\cdots z_{n-2} ;z)\\
= & \sum _{\epsilon ^{\prime } =\pm } \mathsf{C}_{\text{hyp}}\left(\epsilon bp_{0} ,\epsilon^{\prime } bp_{1} ,bs_{1}\right)\tilde{\mathcal{F}}_{\epsilon ^{\prime }}^{[1]}( b;s_{1} ,\cdots ,s_{n-3} ;p_{0} ,\cdots p_{n-1} ;z_{2} ,\cdots z_{n-2} ;z) .
\end{aligned}
\label{eq:degfusion}
\end{equation}

In the semiclassical limit, where $b\rightarrow 0$, $p_{i}\rightarrow \infty $, and $bp_{i} =\theta _{i}$ with $\theta _{i}$ being a finite number, the operator $\phi_{(2,1)}$ behaves as a light operator compared to $\phi\cdots$. Therefore, the correlator $\langle \phi_{(2,1)}(z)\phi \cdots \rangle $ can be interpreted as a light operator $\phi_{(2,1)}$ living in a background fixed by the heavy operators. Thus, we have: 
\begin{align}
 & \langle \phi _{( 2,1)}( z) \phi_{p_{n-1}}( z_{n-1}) \phi_{p_{n-2}}( z_{n-2}) \cdots \phi _{p_{1}}( z_{1}) \phi _{p_{0}}( z_{0}) \rangle\nonumber\\
= & \sum _{s_{1} ,\cdots s_{n-3}} \psi _{s_{1} ,\cdots s_{n-3}}( z) C_{p_{n-1} p_{n-2}}^{s_{n-3}} C_{s_{n-3} p_{n-3}}^{s_{n-4}} \cdots C_{s_{2} p_{2}}^{s_{1}} C_{s_{1} p_{1}}^{p_{0}}\times\nonumber\\[-8pt]
&\hspace{0.35\linewidth}\times\mathcal{F}( b;s_{1} ,\cdots ,s_{n-3} ;p_{0} ,\cdots p_{n-1} ;z_{2} ,\cdots z_{n-2}) ,
\end{align}
where $\Psi _{s_{1} ,\cdots s_{n-3}} (z)$ represents the expectation value of $\phi _{(2,1)} (z)$ in the presence of heavy operators. The conformal block $\mathcal{F}$ in the semiclassical limit exhibits an exponential expression \cite{Zamolodchikov:1987avt,zamolodchikov1986TwodimensionalConformalSymmetry}: 
\begin{equation}
\mathcal{F}( b;s_{1} ,\cdots ,s_{n-3} ;p_{0} ,\cdots p_{n-1} ;z_{2} ,\cdots z_{n-2}) \approx\exp\left(\frac{1}{b^{2}} F( \sigma _{1} ,\cdots \sigma _{n-3} ;\theta _{0} ,\cdots ;z_{2} ,\cdots z_{n-2})\right) ,
\end{equation}
where $\sigma _{i} =bs_{i}$ and $\theta _{i} =bp_{i}$.
This conjecture has been tested in various ways, and a physics-based heuristic explanation is from the saddle point calculation of the path integral in Liouville gravity; see \cite{Harlow:2011ny} for a detailed review of Liouville theory. A proof was later provided in \cite{Besken:2019jyw} using oscillator formalism. By plugging the semiclassical correlator into the equation, we obtain a second-order Fuchsian ODE: 
\begin{equation}
\left(\frac{\partial ^{2}}{\partial z^{2}} +\sum _{k=0}^{n-2}\frac{\delta _{k}}{( z-z_{k})^{2}} +\frac{\delta _{n-1} -\delta _{(1,2)} -\sum _{k=0}^{n-2} \delta _{k}}{z(z-1)} +\sum _{k=2}^{n-2}\frac{( z_{k} -1) \varepsilon _{k}}{z(z-1)( z-z_{k})}\right) \psi(z) =0,
\end{equation}
where $\delta _{k} =\frac{1}{4} -\theta _{k}^{2}$ and $\varepsilon _{k} =z_{k}\frac{\partial F}{\partial z_{k}}$ are the accessory parameters. In the semiclassical limit $b\rightarrow 0$, the semiclassical blocks after fusion can be expanded near their original momentum: 
\begin{equation}
F( \cdots ,\theta+\epsilon b^2/2)  =F( \cdots ,\theta) +\frac{\epsilon b^2}{2}\frac{\partial F}{\partial \theta} +O(b^{4}),
\end{equation}
indicating that
\begin{align}
\tilde{\mathcal{F}}_{\epsilon }^{[0]}( s_{1} ,\cdots ,s_{n-3} ;p_{0} ,\cdots p_{n-1} ;z_{2} ,\cdots z_{n-2} ;z) & =\psi _{\epsilon }^{[ 0]}( z)\exp\left(\frac{\epsilon }{2}\frac{\partial F}{\partial \theta _{0}}\right)\exp\left(\frac{1}{b^{2}} F\right)\label{eq:semidegs} ,\\
\tilde{\mathcal{F}}_{\epsilon ^{\prime }}^{[1]}( s_{1} ,\cdots ,s_{n-3} ;p_{0} ,\cdots p_{n-1} ;z_{2} ,\cdots z_{n-2} ;z) & =\psi _{\epsilon ^{\prime }}^{[ 1]}( z)\exp\left(\frac{\epsilon ^{\prime }}{2}\frac{\partial F}{\partial \theta _{1}}\right)\exp\left(\frac{1}{b^{2}} F\right) .\label{eq:semidegt}
\end{align}
By combining \eqref{eq:degfusion},~\eqref{eq:semidegs},~\eqref{eq:semidegt}, we obtain the connection formula for a Fuchsian equation
\begin{equation}
\psi _{\epsilon }^{[ 0]}( z) =\sum _{\epsilon ^{\prime } =\pm }\mathsf{C}_{\text{hyp}}\left(\epsilon \theta_{0} ,\epsilon^{\prime } \theta_{1} ,\sigma_{1}\right)\exp\left(\frac{\epsilon ^{\prime }}{2}\frac{\partial F}{\partial \theta _{1}} -\frac{\epsilon }{2}\frac{\partial F}{\partial \theta _{0}}\right) \psi_{\epsilon ^{\prime }}^{[1]}(z).\label{eq:Trieste}
\end{equation}

This result is extraordinary; however, there are some inconveniences in application. Specifically, when applying the Trieste formula \eqref{eq:Trieste}, determining $\{\sigma _{1} ,\cdots ,\sigma _{n-3}\}$ requires solving the following set of transcendental equations
\begin{equation}
\varepsilon _{k}( \sigma _{1} ,\cdots ,\sigma _{n-3}) =z_{k}\frac{\partial F( \sigma _{1} ,\cdots ,\sigma _{n-3})}{\partial z_{k}} ,\qquad k=2,\cdots ,n-2.
\end{equation}
In the case of the Heun equation, the coefficient $\sigma _{1}$ can be expanded as a series in $1/z_{2}$, and the coefficients of this series can be analytically derived order by order \cite{Lisovyy:2022flm}. This expansion corresponds to the results in section~\ref{sec:confps}, where we denote $z_{2} =t$. 
\begin{table}[htbp]
\centering
    \begin{footnotesize}
    \begin{tblr}
    {
    hlines,vlines,
    columns = {valign=m,co=-1},
    rows    = {halign=c},
    rowsep=9pt,
    }
    $\mathsf{C}\left(\theta _{0},\theta _{1}\right)$ & $s$-channel & $t$-channel \\
    \SetCell[r=2]{c} Trieste formula  &
    \begin{minipage}[m]{0.4\textwidth}
\begin{tikzpicture}[x=0.75pt,y=0.75pt,yscale=1,xscale=1]
\draw (60,0) -- (120,0) ;
\draw (30,-60) -- (60,0) -- (30,60) ;
\draw (150,-60) -- (120,0) -- (150,60);
\draw[dashed, ->, >=Stealth] (165,0) -- (135,30);
\draw[dashed, ->, >=Stealth] (165,0) -- (135,-30);

\draw (150,-60) node [anchor=west] {$\phi_{p_{0}}$};
\draw (150,60) node [anchor=west] {$\phi_{p_{1}}$};
\draw (30,60) node [anchor=east] {$\phi_{p_{t}}$};
\draw (30,-60) node [anchor=east] {$\phi_{p_{\infty}}$};
\draw (165,0) node [anchor=west] {$\phi_{(2,1)}$};
\draw (90,0) node [anchor=south] {$\phi_{s}$};
\end{tikzpicture}
    \end{minipage}
    & 
    \begin{minipage}[m]{0.4\textwidth}
\begin{tikzpicture}[x=0.75pt,y=0.75pt,yscale=1,xscale=1]
\draw (0,60) -- (0,120) ;
\draw (-60,30) -- (0,60) -- (60,30);
\draw (-60,150) -- (0,120) -- (60,150);
\draw[dashed, ->, >=Stealth] (75,90) -- (30,135);
\draw[dashed, ->, >=Stealth] (75,90) -- (0,90);
\draw[dashed, ->, >=Stealth] (75,90) -- (30,45);

\draw (60,30) node [anchor=west] {$\phi_{p_{0}}$};
\draw (60,150) node [anchor=west] {$\phi_{p_{1}}$};
\draw (-60,150) node [anchor=east] {$\phi_{p_{t}}$};
\draw (-60,30) node [anchor=east] {$\phi_{p_{\infty}}$};
\draw (0,90) node [anchor=east] {$\phi_{s}$};
\draw (75,90) node [anchor=west] {$\phi_{(2,1)}$};
\end{tikzpicture}
    \end{minipage}
    \\
     & $\displaystyle \mathsf{C}_{\text{hyp}}\left(\theta_{0} ,\theta_{1}, \sigma\right)\exp\left(\frac{1}{2}\frac{\partial F}{\partial \theta _{1}} -\frac{1}{2}\frac{\partial F}{\partial \theta _{0}}\right) $ & 
    \begin{minipage}[m]{0.4\textwidth}
    $\begin{aligned}
 & \sum _{\kappa =\pm }\mathsf{C}_{\text{hyp}}( \theta _{0} ,\kappa \sigma ,\theta _{\infty })\mathsf{C}_{\text{hyp}}(-\kappa \sigma ,\theta _{1} ,\theta _{t}) \times \\
 & \ \ \ \ \ \ \ \ \times \exp\left(\frac{\kappa }{2}\frac{\partial F}{\partial \sigma } +\frac{1}{2}\frac{\partial F}{\partial \theta _{1}} -\frac{1}{2}\frac{\partial F}{\partial \theta _{0}}\right)
\end{aligned}$
    \end{minipage}
    \\
    Recurrence relation & $\displaystyle  \mathsf{C}_{\text{hyp}}\left(\theta_{0} ,\theta_{1}, \mathsf{w}\right)\left(1-\frac{1}{t}\right)^{1/2 -\theta _{t}}a_{\infty}$ & To be found \\
    Series expansion & around $t=\infty$ & around $t=1$ \\
    \end{tblr}
    \end{footnotesize}
    \caption{\label{tab:stchannel}Comparison between the Trieste formula and the recurrence relation in $s$ and $t$ channels. A solid line represents a primary operator, and a dashed line with an arrow represents the degenerate operator and its possible insertion position.}
\end{table}

Using the crossing symmetry of the conformal blocks, a series expansion can also be obtained for the connection formula around $t=1$, as provided by the $t$-channel semiclassical block and shown in table~\ref{tab:stchannel}. In the $s$-channel expansion, $\phi _{p_{0}}$ and $\phi _{p_{1}}$ are directly expanded using OPE, allowing the connection formula to be obtained via a single fusion transformation, as previously discussed. In the $t$-channel expansion, $\phi _{p_{0}}$ and $\phi _{p_{1}}$ are not directly expanded using OPE. However, the insertion of a degenerate operator enables fusion with $\phi _{s}$, and the connection formula is then derived through two sequential fusion transformations. See a detailed discussion of the $t$-channel Trieste formula in \cite{Jia:2024zes}. Therefore, we believe that a connection formula based on recurrence relation corresponding to the $t$-channel expansion exists but has not yet been found. Moreover, the Trieste formula also contains the confluent cases of the Heun equation, expressed by irregular conformal blocks \cite{Bonelli:2022ten,Bonelli:2021uvf}. An important scenario involves the connection between a regular singularity and an irregular singularity. Although a version of the connection formula based on a recurrence relation exists \cite{schäfke1984ConnectionProblemRegular}, it requires further refinement to provide an analytic series expansion.

\section{Discussion}
\label{sec:sum}

We have taken two physics examples to show how to take advantage of the recent advances in the connection formula for the Heun function to express the holographic correlators, if the perturbation equation of a black hole has four regular singularities. The main conclusions are summarized as follows.

\begin{itemize}
\item We express the thermal correlators in terms of a recurrence relation. 

\item We reexamined the RN-AdS$_5$ black hole perturbed by a charged scalar field, giving a Heun equation as a nontrivial example to test the recursion formula.

\item We examined a new example of another charged black hole that corresponds to a different state in $\mathcal{N}=4$ SYM theory.

\item We briefly review the Triest formula, which uses the Virasoro conformal blocks to express the connection formula. The recursion formula is equivalent, but significantly simpler and easier to implement.
\end{itemize}

The following topics need further investigation:
\begin{itemize}
    \item The recurrence relation \eqref{eq:recur} corresponds to the $\lambda=1/t$ expansion of the perturbative connection formula. What is the recurrence relation corresponding to the $\lambda=t-1$ expansion of the perturbative connection formula?
    \item Either when the temperature approaches zero, or when the cosmological constant approaches zero, we obtain irregular singularities. The extremal limit of the charged AdS black holes gives a confluent Heun equation, which has two regular singularities and one irregular singularity of rank 1. What is the recurrence relation in this case?

    \item Applying to other black hole perturbations. The recursion approach is applicable as long as the perturbation equation can be transformed to an ODE with four regular singularities, two of which have boundary conditions being imposed.
\end{itemize}

\acknowledgments
We thank Oleg Lisovyy for helpful communications. This work was supported in part by the NSF of China under Grant No. 11905298.

\appendix

\section{Universal and Schäfke-Schmidt connection formulas}
\label{sec:conn}

There is a universal connection formula for a second-order ODE with given linearly independent solutions $\{\psi_1^{[0]}, \psi_2^{[0]}\}$ around $z=0$ and $\{\psi_1^{[1]}, \psi_2^{[1]}\}$ around $z=1$:
\begin{equation}
    \begin{pmatrix}
         \psi_1^{[0]}(z)\\
         \psi_2^{[0]}(z) 
    \end{pmatrix}=
    \frac{1}{W[\psi_1^{[1]},\psi_2^{[1]}]}
    \setlength\arraycolsep{5pt}
    \begin{pmatrix}
         W[\psi_1^{[0]},\psi_2^{[1]}] & -W[\psi_1^{[0]},\psi_1^{[1]}]\\
         W[\psi_2^{[0]},\psi_2^{[1]}] & -W[\psi_2^{[0]},\psi_1^{[1]}]
    \end{pmatrix}
    \begin{pmatrix}
         \psi_1^{[1]}(z)\\
         \psi_2^{[1]}(z) 
    \end{pmatrix},
\end{equation}
where $W$ is the Wronskian, which is independent of $z$. In practice, we can take a middle value $z=1/2$ to evaluate it numerically. Although this formula is useful to check the correctness of a connection formula, it is usually inefficient to evaluate the local solutions and their Wronskians. It is desirable to obtain a connection formula in terms of simpler functions. For example, the connection formula for the hypergeometric equation can be written in terms of gamma functions.

In \cite{schäfke1980ConnectionProblemGeneral, schäfke1980ConnectionProblemTwo}, R. Schäfke and D. Schmidt studied the connection problem for a linear ODE at two regular singular points. The recurrence relation \eqref{eq:recur} is an improved version of the Schäfke-Schmidt connection formula, which is
\begin{equation}
\mathsf{C}( \theta _{0} ,\theta _{1}) =\Gamma ( 2\theta _{1})\lim _{k\rightarrow \infty } k^{1-2\theta _{1}} u_{k} \label{eq:genpsiconn}
\end{equation}
for \eqref{eq:gen2ndODE}, according to Corollary 3.2 in \cite{Lisovyy:2022flm}. Here the coefficients $u_k$ are defined by
\begin{equation}
z^{-\frac{1}{2} +\theta _{0}}( 1-z)^{-\frac{1}{2} -\theta _{1}} \psi _{+}^{[0]} (z)=1+\sum _{k=1}^{\infty } u_{k} z^{k}.
\end{equation}
In the case of connection between regular singularities of the Heun equation, by defining $a_k=u_k/u_k^{t\to\infty}$ to cancel the explicit dependence of $k^{1-2\theta _{1}}$, the formula can be expressed as an analytic series expansion \cite{Lisovyy:2022flm}. Furthermore, this redefinition improves the efficiency of numerical computations.

\section{Hypergeometric equation}
\label{sec:hypergeom}

A comparison between the hypergeometric equation and the Heun equation is helpful in understanding their connection formulas. We include both the standard form and normal form of the equation.

The standard form of the hypergeometric equation is
\begin{equation}
    z(1-z)y''(z)+\left(c-(a+b+1)z\right)y'(z)-aby(z)=0\,.
\end{equation}
The three regular singular points $\{0,1,\infty\}$ and the corresponding exponents are represented by the Riemann scheme as
\begin{equation}
    y=P
    \setlength\arraycolsep{5pt}
    \begin{Bmatrix}
    0 & 1 & \infty &\\
    0 & 0 & a & ;z\\
    1-c & c-a-b & b &
    \end{Bmatrix}.
\end{equation}
The two linearly independent solutions near $z=0$ are
\begin{align}
    y^{[0]}_1 &={_2F_1}(a,b;c;z)\,,\\
    y^{[0]}_2 &=z^{1-c}\,{_2F_1}(a - c + 1, b - c + 1; 2 - c; z)\,.
\end{align}
The two linearly independent solutions near $z=1$ are
\begin{align}
    y^{[1]}_1 &={_2F_1}(a, b; a + b - c + 1; 1 - z)\,,\\
    y^{[1]}_2 &=z^{c-a-b}\,{_2F_1}(c - a, c - b; c - a - b + 1; 1 - z)\,.
\end{align}
The two sets of solutions are connected by
\begin{equation}
\begin{pmatrix}
    y^{[0]}_1\\
    y^{[0]}_2
\end{pmatrix}=
\setlength\arraycolsep{5pt}
\begin{pmatrix}
 \frac{\Gamma (c) \Gamma (c-a-b)}{\Gamma (c-a) \Gamma (c-b)} & \frac{\Gamma (a+b-c) \Gamma (c)}{\Gamma (a) \Gamma (b)} \\
 \frac{\Gamma (2-c) \Gamma (c-a-b)}{\Gamma (1-a) \Gamma (1-b)} & \frac{\Gamma (2-c) \Gamma (a+b-c)}{\Gamma (a-c+1) \Gamma (b-c+1)}
\end{pmatrix}
\begin{pmatrix}
    y^{[1]}_1\\
    y^{[1]}_2
\end{pmatrix}.
\end{equation}
Boundary value problems between $z=0$ and $z=1$ can be analytically solved by means of the connection formula.

The normal form of the hypergeometric equation is
\begin{equation}
   \left(\frac{d^2}{dz^2}+\frac{\frac14-\theta_0^2}{z^2}+\frac{\frac14-\theta_1^2}{(z-1)^2}+\frac{\theta_0^2+\theta_1^2-\theta_\infty^2-\tfrac14}{z(z-1)}\right)\psi(z)=0\,,
\end{equation}
which is related to the standard form by $\psi(z)=z^{1/2-\theta_0}(1-z)^{1/2-\theta_1}y(z)$ and
\begin{equation}
    \theta_0=\frac{1}{2}(1-c),\qquad \theta_1=\frac{1}{2}(c-a-b),\qquad \theta_\infty=\frac{1}{2}(b-a)\,.
\end{equation}
The three regular singular points and the corresponding exponents are represented by the Riemann scheme as
\begin{equation}
    \psi=P
    \setlength\arraycolsep{5pt}
    \begin{Bmatrix}
    0 & 1 & \infty &\\
    \frac{1}{2}-\theta_0 & \frac{1}{2}-\theta_1 & \frac{1}{2}-\theta_\infty & ;z\\
    \frac{1}{2}+\theta_0 & \frac{1}{2}+\theta_1 & \frac{1}{2}+\theta_\infty &
    \end{Bmatrix}.
\end{equation}
The two linearly independent solutions near $z=0$ are
\begin{equation}
    \psi^{[0]}_{\pm}=z^{1/2\mp\theta_0}(1-z)^{\frac12-\theta_1}\,{_2F_1}(\tfrac12\mp\theta_0-\theta_1-\theta_\infty,\tfrac12\mp\theta_0-\theta_1+\theta_\infty;1\mp 2\theta_0;z)\,.
\end{equation}
The two linearly independent solutions near $z=1$ are
\begin{equation}
    \psi^{[1]}_{\pm}=(1-z)^{\frac12\mp\theta_1}z^{1/2-\theta_0}\,{_2F_1}(\tfrac12-\theta_0\mp\theta_1-\theta_\infty,\tfrac12-\theta_0\mp\theta_1+\theta_\infty;1\mp 2\theta_1;z)\,.
\end{equation}
The connection formula can be written as
\begin{equation}
    \psi^{[0]}_{\epsilon}=\sum_{\epsilon'}\mathsf{C}_{\text{hyp}}(\epsilon\theta_0,\epsilon'\theta_1,\theta_\infty)\psi^{[1]}_{\epsilon'},\qquad \epsilon,\epsilon'=\pm\,,
\end{equation}
where
\begin{equation}
    \mathsf{C}_{\text{hyp}}(\theta_0,\theta_1,\theta_\infty)=\frac{\Gamma(1-2\theta_0)\Gamma(2\theta_1)}{\Gamma(\frac{1}{2}-\theta_0+\theta_1+\theta_\infty)\Gamma(\frac{1}{2}-\theta_0+\theta_1-\theta_\infty)}\,.
\end{equation}

\section{Comparison with the pseudospectral method}
\label{sec:comp}
\begin{figure}
    \centering
    \includegraphics[height=0.295\linewidth]{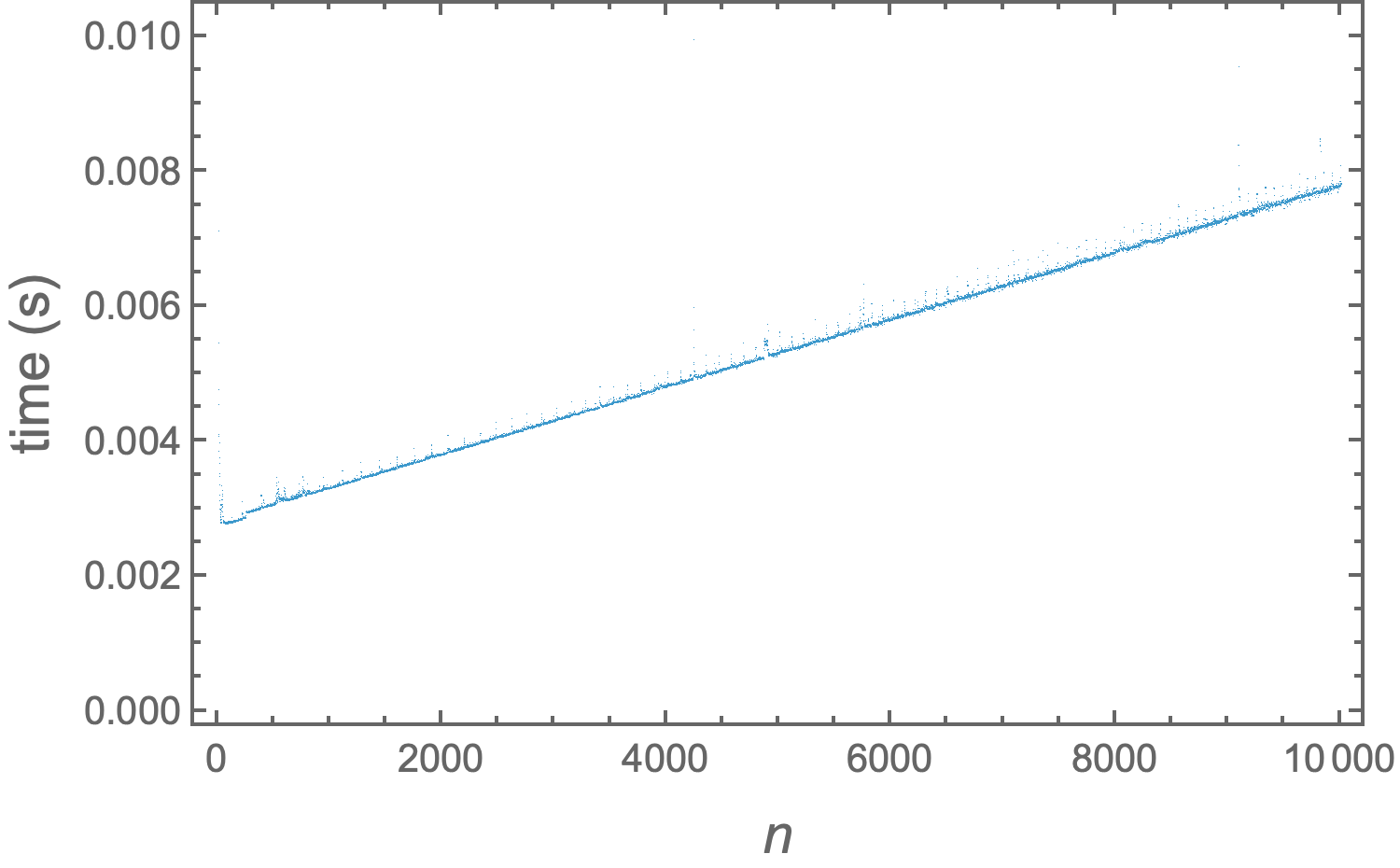}\quad
    \includegraphics[height=0.295\linewidth]{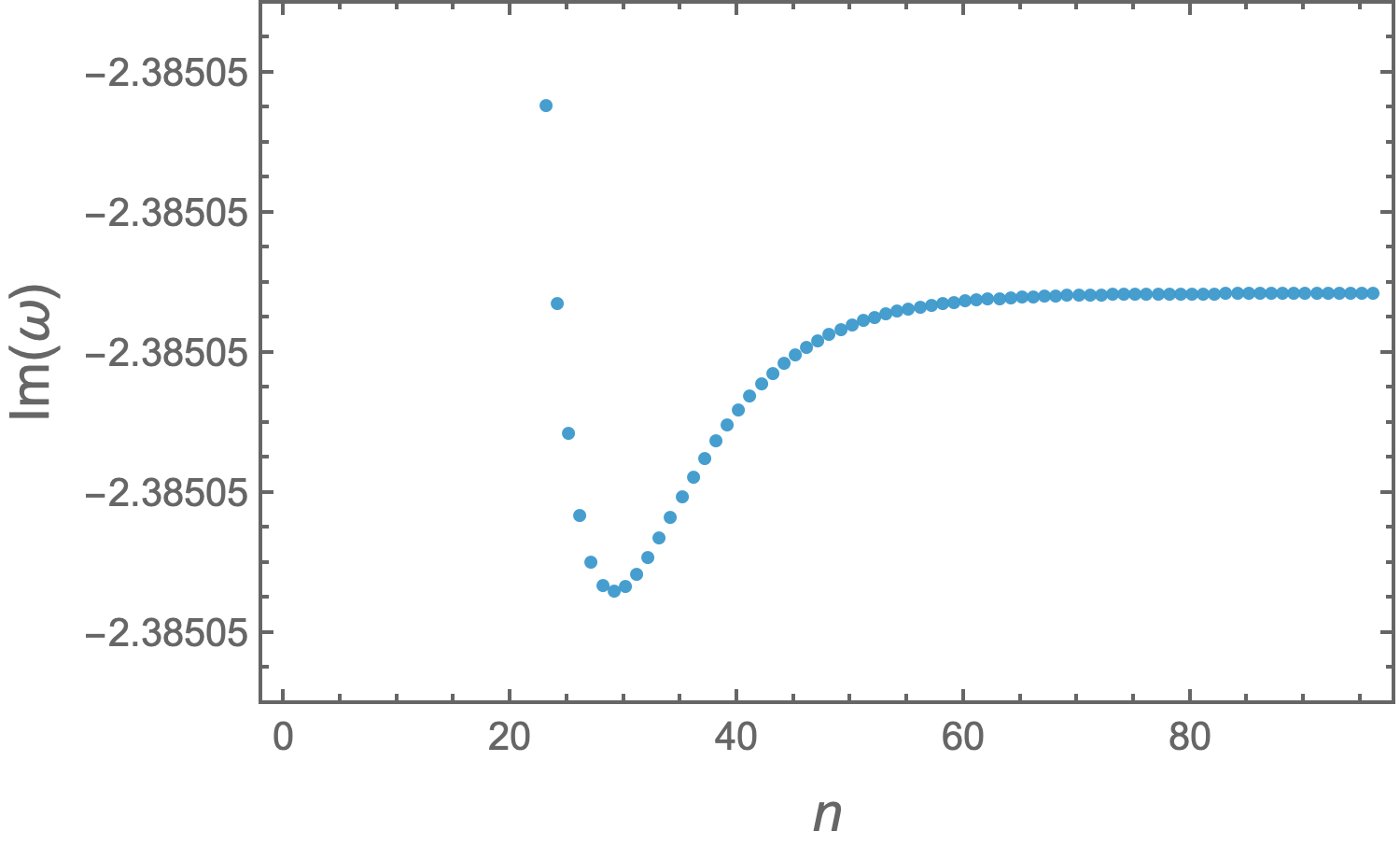}
    \caption{The left panel shows the time to obtain $a_n$ by the recursion relation for a typical set of parameters. The time grows linearly in $n$ with a small slope. It takes roughly twice the time for $n=10^4$ compared with $n=100$. The right panel shows the convergence of the recurrence relation. The imaginary part of the QNM is within $\mathcal{O}(10^{-6})$ accuracy for $n$ less than 100.}
    \label{fig:conv}
\end{figure}

We discuss the convergence, stability, accuracy, and effectiveness of the recurrence relation by comparing quantitatively with QNMs calculated by the pseudospectral method. Firstly, the recurrence relation is efficient to evaluate numerically and converges very fast. The left panel of figure~\ref{fig:conv} shows that the time to obtain the $n$th term $a_n$ grows linearly in $n$ with a small slope. The right panel of figure~\ref{fig:conv} shows that for a typical choice of parameters, the QNM converges with good accuracy within 100 recursions. We compare the QNMs for the Schwarzschild-AdS black hole in an early work \cite{Horowitz:1999jd} and find perfect agreement, as shown in table~\ref{tab:HH}.

\begin{table}[h]
\renewcommand{\arraystretch}{1.2}
    \centering
    \begin{tabular}{|c|c|c|}
    \hline
    $r_+$  & Horowitz \& Hubeny 1999 & Recursion method \\
    \hline
    100 & $311.9627-274.6655i$ & $311.96274 - 274.66547 i$\\
50 & $156.0077-137.3296\,i$ & $156.00775 - 137.32957 i$\\
10 & $31.3699-27.4457i$ & $31.36991 - 27.44570 i$\\
5 & $15.9454-13.6914i$ & $15.94542 - 13.69141 i$\\
1 & $4.5788-2.5547i$ & $4.57877 - 2.55468 i$\\
0.8 & $4.1951-1.9676i$ & $4.19507 - 1.96754 i$\\
0.6 & $3.8914-1.3656i$ & $3.89140 - 1.36561 i$\\
0.4 & $3.7174-0.7462i$ & $3.71739 - 0.74625 i$\\
    \hline
    \end{tabular}
    \caption{The lowest QNMs for the Schwarzschild-AdS black hole with different horizon sizes. The column of Horowitz \& Hubeny 1999 is from table 1 of \cite{Horowitz:1999jd}. The last column is calculated using the recursion method with 100 recursions and machine precision (16 digits).}
    \label{tab:HH}
\end{table}

\begin{table}[h]
\renewcommand{\arraystretch}{1.2}
    \centering
    \begin{tabular}{|c|c|}
    \hline
     Quasinormal modes in figure 1 ($q=0$) & Difference\\
    \hline
     $-2.38504913458487023569 i$ & $\mathcal{O}(10^{-37})$\\
 $\pm 4.69462676089200666955 - 3.57737148481108649718 i$ & $\mathcal{O}(10^{-35})$\\
 $-5.23308505113461701067 i$ & $\mathcal{O}(10^{-31})$\\
 $\pm 7.69044007645390049083 - 6.37557154286504575354 i$ & $\mathcal{O}(10^{-30})$\\
 $-8.34596068752678524732 i$ & $\mathcal{O}(10^{-26})$\\
 $\pm 10.67911520322740826939 - 9.11638217721954816814 i$ & $\mathcal{O}(10^{-26})$\\
 $-11.53098086344363155484 i$ & $\mathcal{O}(10^{-22})$\\
 $\pm 13.66616848147946013990 - 11.83869347250288104955 i$ & $\mathcal{O}(10^{-23})$\\
    \hline
    \end{tabular}
    \caption{\label{tab:my_label} The first 12 QNMs for the RN-AdS$_5$ black hole sorted by their imaginary parts. The parameters are the same as the first plot of figure~1. For the pseudospectral method, we use a grid size of 80 with a precision of 40 digits. For the recursion method, we take $10^3$ recursions with a precision of 40 digits. The difference between the absolute value of the QNMs calculated by the two methods is shown in the second column.}
\end{table}

We compare the quasinormal modes calculated by the pseudospectral method,\footnote{We use a Mathematica package available at \href{https://github.com/renphysics/GRSpectral}{https://github.com/renphysics/GRSpectral}.} which uses up to $N$th-order Chebyshev polynomials as bases to approximate the eigenfunctions. (The grid size $N$ in the following means $N+1$ collocation points.) A major advantage of the pseudospectral method is that no initial guess of the modes is needed since we are solving a linear system. When the temperature is decreased, it is essential to increase the grid size and the precision accordingly. The computational complexity grows significantly as $\mathcal{O}(N^3)$.

For the recursion method, we can increase the number of recursions to $10^3$ or $10^4$ and the precision to 40 digits to calculate the QNMs with larger imaginary parts at lower temperatures. For $n=10^4$, the computation takes only a few times longer than it does for $n=100$. While the two methods agree at a high accuracy, the time for the recursion method grows much more slowly than the pseudospectral method as the temperature is decreased. A complementary way is to use the pseudospectral method to generate the QNMs as an initial guess for the recursion method, and gradually change the parameters by the method of continuation. The details for four sets of parameters are as follows.

\begin{itemize}
    \item $r_+=1$, $Q=1.2$, $q=0$: The dimensionless temperature $T/\mu=0.119$. For the pseudospectral method, we use a grid size $N=200$ and a precision of 100 digits. For the recursion method, we take $n=10^3$ recursions and 40 digits of precision. We find a uniform $\mathcal{O}(10^{-38})$ accuracy for the first 20 QNMs sorting by their absolute value.
    
    \item $r_+=1$, $Q=1.4$, $q=3$: The dimensionless temperature $T/\mu=0.068$. We use the same other parameters as above and find a uniform $\mathcal{O}(10^{-24})$ accuracy for the first 20 QNMs.
    
    \item $r_+=0.9$, $Q=1.2$, $q=1$: The dimensionless temperature $T/\mu=0.029$. For the pseudospectral method, we use a grid size $N=400$ and a precision of 200 digits. For the recursion method, we take $n=10^3$ recursions and 40 digits of precision. We find a uniform $\mathcal{O}(10^{-13})$ accuracy for the first 20 QNMs. Increasing the number of recursions to $10^4$ achieves a uniform $\mathcal{O}(10^{-33})$ accuracy.
    
    \item $r_+=0.9$, $Q=1.28$, $q=1$: The dimensionless temperature $T/\mu=0.008$. For the pseudospectral method, we use grid sizes $N=200$, $400$, and $800$ and a precision of 300 digits. For the recursion method, we take $n=10^4$ recursions and 40 digits of precision. We find a uniform $\mathcal{O}(10^{-3})$, $\mathcal{O}(10^{-15})$, $\mathcal{O}(10^{-30})$ accuracy for the first 20 QNMs.
\end{itemize}

\bibliographystyle{JHEP}
\bibliography{biblio.bib}
\end{document}